\documentclass[letterpaper,twoside,twocolumn,english,reprint,aps,pre,showpacs]{revtex4-1}
\usepackage[T1]{fontenc}
\usepackage[utf8]{inputenc}
\usepackage{textcomp}
\usepackage{amsbsy}
\usepackage{amstext}
\usepackage{graphicx}
\usepackage{esint}

\makeatletter

\pdfpageheight\paperheight
\pdfpagewidth\paperwidth

\providecommand{\tabularnewline}{\\}

 \@ifundefined{textcolor}{}
 {%
   \definecolor{BLACK}{gray}{0}
   \definecolor{WHITE}{gray}{1}
   \definecolor{RED}{rgb}{1,0,0}
   \definecolor{GREEN}{rgb}{0,1,0}
   \definecolor{BLUE}{rgb}{0,0,1}
   \definecolor{CYAN}{cmyk}{1,0,0,0}
   \definecolor{MAGENTA}{cmyk}{0,1,0,0}
   \definecolor{YELLOW}{cmyk}{0,0,1,0}
 }

\usepackage{babel}

\makeatother

\usepackage{babel}
\begin{document}

\title{Shallow granular flows down flat frictional channels: steady flows
and longitudinal vortices.}

\author{Nicolas Brodu}

\author{Patrick Richard}

\author{Renaud Delannay}

\affiliation{Institut Physique de Rennes, UMR CNRS 6251, Université de Rennes
1, Campus de Beaulieu Bâtiment 11A, 263 av. Général Leclerc, 35042
Rennes CEDEX}
\begin{abstract}
Granular flows down inclined channels with smooth boundaries are common
in nature and industry. Nevertheless, flat boundaries have been much
less investigated than bumpy ones, which are used by most experimental
and numerical studies to avoid sliding effects. Using DEM numerical
simulations with side walls we recover quantitatively experimental
results. At larger angles we predict a rich behavior, including granular
convection and inverted density profiles suggesting a Rayleigh-Bénard
type of instability. In many aspects flows on a flat base can be seen
as flows over an effective bumpy base made of the basal rolling layer,
giving Bagnold-type profiles in the overburden. We have tested a simple
viscoplastic rheological model (Nature 2006, vol 441, pp727-730) in
average form. The transition between the unidirectional and the convective
flows is then clearly apparent as a discontinuity in the constitutive
relation. 
\end{abstract}


\maketitle

\section{Introduction}

Granular dense flows down inclined channels preserve the complexity
of granular flows while remaining simple enough for a detailed analysis
\cite{Delannay2007}. They are of interest in engineering applications
involving conveying of solid materials such as minerals, or in geophysical
situations like rock avalanches or pyroclastic flows. This article
focuses on flows down a flat, frictional incline. These flows differ
substantially from those on a rough or bumpy base with macroscopic
asperities on the order of the diameter of the flowing particles.
We have developed simulations that model the experimental configuration
used by Louge and Keast \cite{Louge_Keast_2001}: shallow flows ($\approx7$
grain diameters at rest height, see Fig. \ref{fig:Scenario-description})
in a wide ($\approx68$ grains) chute with flat frictional surfaces.
We investigate flows in a range of inclination angles containing the
range of experimentally observed Steady and Fully Developed (SFD)
flows. We reproduce the flow properties quantitatively and analyze
the internal flow structure. We show that above a given inclination
angle granular convection occurs in association with inverted density
profiles. To our best knowledge our work is the first to predict that
secondary flows also exist with flat boundaries for SFD flows. The
basal rolling layer can be seen as an effective ``bumpy'' base for
the core flow sliding on top of it \cite{Delannay2007,Taberlet_2005}.
In these conditions, we show that velocities in the main bulk of the
flow follow a Bagnold scaling. This type of flows is associated with
a constant homogeneous inertial number for SFD flows on a bumpy base
\cite{GDRMIDI2004}. This led us to study the rheology of these flows,
and test whether the viscoplastic rheology holds.

The paper is structured as follows. The next section, which can be
skipped by specialists of the field, is devoted to the state of the
art. Section 3 gives details about the simulation method we will use.
Global properties of the flows are studied in section 4. Section 5
is devoted to detailed results concerning the packing fraction, pressure,
velocity and ``granular temperature'' fields. The rheological study
is presented in Section 6. Concluding remarks are given in Section
7.

\section{The state of the art on granular flows down inclined channels}

\label{sec:state}Significant progress have been made during the last
decades in describing dense granular flows, nevertheless they continue
to resist our understanding and remain an active field of research.
Very dense quasistatic regimes are usually described by plastic models
\cite{Nedderman_Laohakul_1980}, and a kinetic theory of granular
gas has been developed \cite{Jenkins_Savage_1983} that can accurately
render the behavior of dilute flows. A viscoplastic description for
dense fluid regimes has been proposed based on a dimensionality analysis
in the unidirectional case \cite{GDRMIDI2004}. This rheology has
then been extended to 3D for incompressible flows \cite{Jop_etal_2006}.
In all these cases, when the parameters are set according to the expected
theoretical values (\textit{e.g.} high packing fraction dense uniform
flows, initial and boundary conditions, etc.), the proposed constitutive
equations match the experiment (\textit{e.g.} collapse of velocity
profiles \cite{Pouliquen1999}). Extensions are then proposed to account
for variations of nearly related cases: an extended granular gas theory
taking into account correlations in denser cases \cite{JenkinsBerzi2010},
or a variant of the viscoplastic model for compressible flows \cite{Borzsonyi_etal_2009}.
Despite all these efforts, a comprehensive theory is still missing
in the general and most common case where the coexistence of both
dense and dilute parts are observed within the same flow, and which
would correctly incorporate the influence of boundary conditions such
as sidewalls and bottom.

A large corpus of studies exists on dense granular flows down an inclined
plane chute (more than 100 references in \cite{GDRMIDI2004}, additional
ones in \cite{Delannay2007}). The boundaries are known to change
the flow structure \cite{Savage1979}. The choice of wide channels
is an attempt to avoid the influence of sidewalls. In the same way,
numerical simulations in periodic cells attempt to study flows down
infinitely long and wide chutes. Most experimental and numerical works
avoid the inherent discontinuity and sliding at the base by covering
the surface with glued, fixed grains of the same nature as these involved
in the flow \cite{Savage1979,Pouliquen1999,Silbert_etal_2001,Forterre_Pouliquen_2002}.
In these conditions there exist limits on the lower inclination angle
and the piling height below which the grains do not flow. Above these
thresholds and for moderate inclinations, dense fluid flows present
a negligible velocity at the bumpy base. Thin SFD flows comprising
a few layers of grains exhibit a nearly linear, sheared vertical profile
of the velocity (\cite{GDRMIDI2004}, 2D experiments \cite{Bi2006,Azanza_1997,Berton_etal_2003},
2D \cite{Dippel_1998} and 3D periodic \cite{Silbert_etal_2001} numerical
simulations). For thicker flows, a Bagnold scaling is observed in
the core of the SFD flow, with lower velocities at the base (\cite{GDRMIDI2004},
3D periodic numerical simulations \cite{Silbert_etal_2001}). At larger
angles of inclination the flows are more dilute and an inverted density
profile is observed \cite{Taberlet2007,Borzsonyi_etal_2009,Forterre_Pouliquen_2001}.
This inversion was analyzed by means of the granular gas theory to
induce a Rayleigh-Bénard type of instability \cite{Forterre_Pouliquen_2001,Forterre_Pouliquen_2002}.
The convection rolls take the form of longitudinal stripe patterns
\cite{Forterre_Pouliquen_2002,Borzsonyi_etal_2009}. These can be
reproduced numerically using Periodic Boundary Conditions (PBC) \cite{Borzsonyi_etal_2009}
provided the width the periodic cell, $W$, is large enough compared
to the grain diameter $D$ for the convection rolls to appear. Convection
has never been observed in numerical works using $W=10D$ \cite{Silbert_etal_2001,Silbert2002}.
Below about $W\approx50D$, there seems to be not enough space for
developing convection rolls \cite{Borzsonyi_etal_2009}.

Beside these studies of ``unconfined'' flows in large channels,
extensive measurements highlighting the influence of walls were performed~\cite{Ancey2001,Taberlet2003,Bi2005,Jop_etal_2005}.
Experimental and numerical studies both in 2D and 3D configurations
\cite{Bi2005} show that frictional lateral walls alter the flow properties.
For instance, SFD flows on bumpy bases are observed up to large inclination
angles where accelerated ones are usually expected. Moreover, at any
given inclination angle, there is a critical flow rate above which
a static heap forms along the base. The heap is stabilized by the
flow atop it \cite{Taberlet2003}. Flows atop this sidewall-stabilized
heap (SSH) differ from SFD flows on bumpy base as they occur over
erodible bases, but still display SFD features. The effect of side
walls on SFD flows on top of a static pile in a channel has been studied
by carrying out experiments in setup of different widths, up to 600
particle diameters \cite{Jop_etal_2005}. They show that these flows
are entirely controlled by side wall effects.

\begin{figure}
\begin{centering}
\includegraphics[width=1\columnwidth]{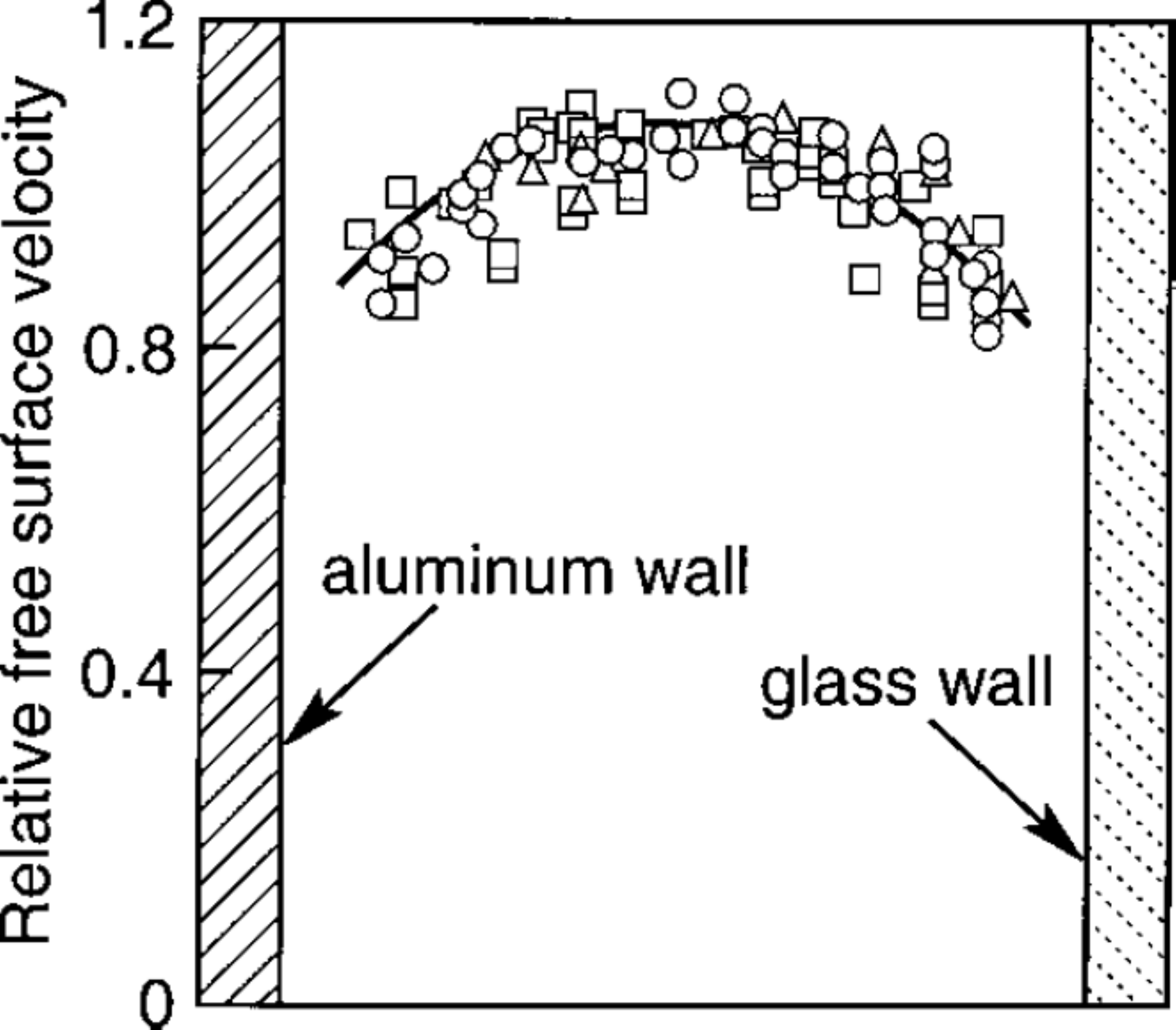}
\par\end{centering}

\caption{\label{fig:Louge_Keast_2001_reprint}Top view of free surface velocity
profiles for granular flows down flat and frictional base (reprinted
from~\cite{Louge_Keast_2001}). The influence of the side walls is
apparent over 2/3 of the flow width. }
\end{figure}

The bumpy bases made of glued grains case is thus relatively well-studied.
However most industrial conditions involve flat boundaries, as well
as natural flows occurring on smooth bed rocks at the scale of the
grains. Surprisingly very few studies \cite{Savage1979,Campbell_Brennen_1985,Campbell_etal_1985,Patton_etal_1987,Johnson_etal_1990,Ahn_etal_1991,Ahn_etal_1992,Walton1993,Sadjadpour_Campbell_1999,Louge_Keast_2001}
have considered this more common case of flat frictional surfaces.
Early experimental works mention increased flow velocity and sliding
conditions at the boundaries compared to the bumpy walls case~\cite{Savage1979}.
Differences with the bumpy case situation are manifest in the flow
properties. Velocity profiles (transversal and in height) involve
a slip condition at the boundaries \cite{Ahn_etal_1991}, compared
to the null velocity condition at the interface with the bumpy walls.
Some unexplained surging waves are occasionally observed at the surface
\cite{Savage1979,Louge_Keast_2001}, blurring its exact location by
a layer of grains in saltation. More recently Louge and Keast \cite{Louge_Keast_2001}
conducted experiments on a flat base with a well documented set of
parameters, with a more detailed analysis than previous experimental
works on the topic \cite{Ahn_etal_1991,Ahn_etal_1992,Patton_etal_1987,Johnson_etal_1990}.
They confirmed the aforementioned observations regarding the flow
structure and velocity profiles. A layer of rolling grains with intermittent
jumps develops on the flat base, with the rest of the flow sliding
on top of this basal layer. The influence of the distant side-walls
is negligible in terms of induced friction: ``the relative contribution
of side walls in the force balance {[}…{]} never exceeds 7\%'' \cite{Louge_Keast_2001}.
However this influence of the side walls is clearly apparent over
2/3 of the flow width (Fig.~\ref{fig:Louge_Keast_2001_reprint}),
reprinted from \cite{Louge_Keast_2001}), leading to the conclusion
that some other mechanism is involved for a long-range influence of
the boundary conditions. 

\begin{figure}
\begin{centering}
\includegraphics[width=1\columnwidth]{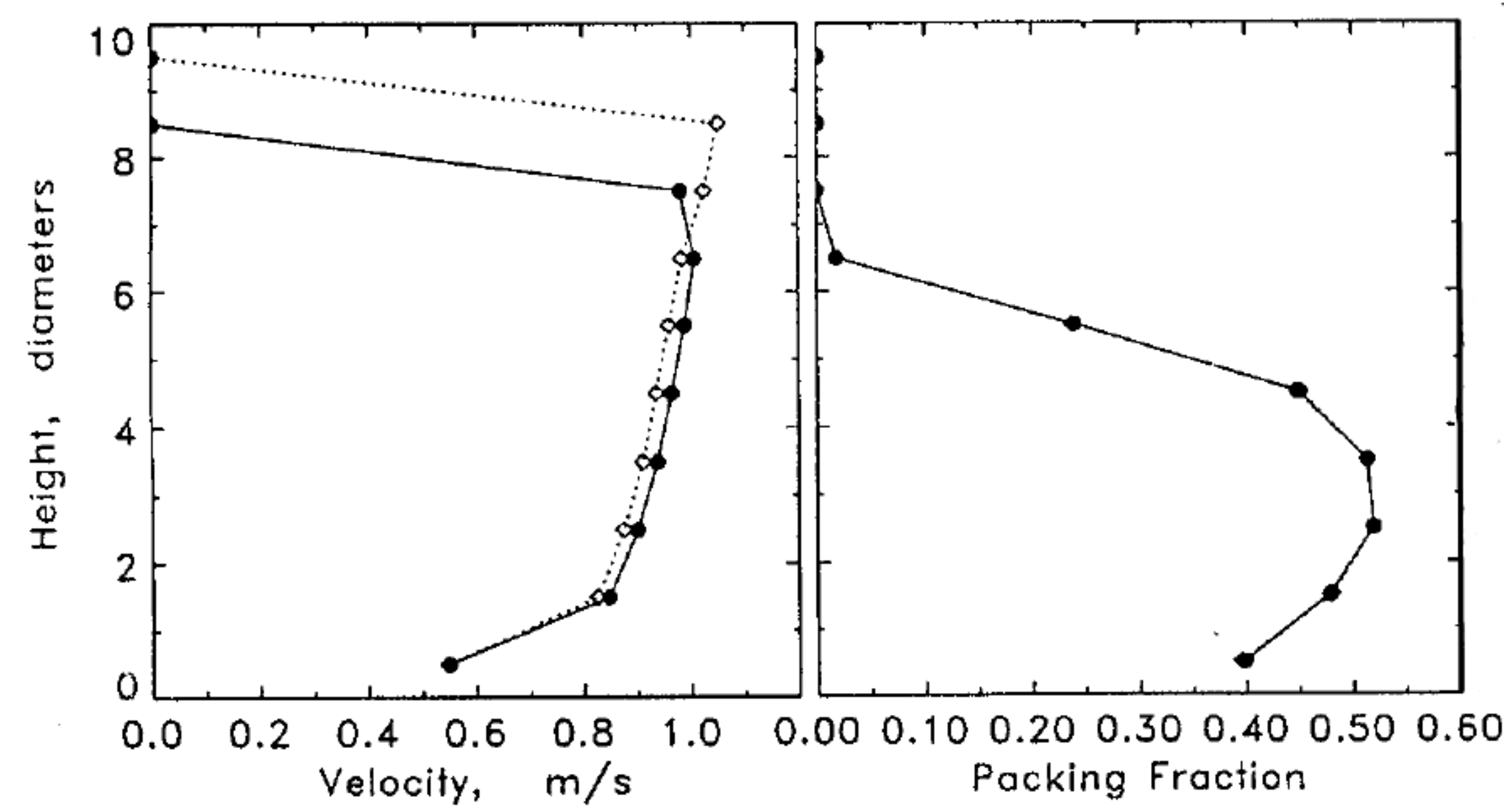}
\par\end{centering}
\caption{\label{fig:Walton}. Velocity (left) and packing fraction (right)
profiles obtained by Walton~\cite{Walton1993} for moderate flow
heights. Solid lines are instantaneous profiles, dotted line is time
average profiles. Reprinted from~\cite{Walton1993}.}
\end{figure}

They also reported a range of inclination angles $\left[\theta_{\textrm{min}},\theta_{\textrm{max}}\right]$
for the observation of steady fully developed (SFD) flows, independent
of the flow height, that presents a much lower bound than in the bumpy
surface case \cite{Pouliquen1999}. The upper bound for SFD flows
is also provided, but as correctly pointed out by \cite{Haynes_Walton_2000}
the attainment of SFD flows is restricted by the physical length of
the chute, hence so is the maximal angle above which an ``accelerated
regime'' is observed. Numerical simulations can be used in order
to complement these experimental results, but literature on numerical
studies over flat frictional surfaces is sparse. Early simulations
are reported in \cite{Campbell_Brennen_1985} (2D) and \cite{Walton1993}
(3D). Given their limited computational power, implying the use of
a small periodic cell and a low number of grains, and given their
use of monodisperse grains, a direct comparison with the experiment
is difficult. More recent numerical works considering a flat base~\cite{Taberlet_2005}
do not provide a detailed analysis of its influence. Without sidewalls
SFD flows on a flat base can only be sustained for inclination angles
whose tangent is less than the friction coefficient \cite{Louge_Keast_2001,Walton1993}.
Thus, with PBC, the maximal inclination angle $\theta_{\textrm{max}}$
is fixed by the solid friction on the base: $\mu=\tan\theta_{\textrm{max}}$.
Walton \cite{Walton1993} got effectively SFD flows for inclination
angles $\theta$ whose tangent is smaller than the friction coefficient
$\mu$, else they accelerate unboundedly. Nevertheless the experimental
value of $\theta_{\textrm{max}}$ lead to choosing a large value of
$\mu$ which is not compatible with the friction coefficients measured
in impacts \cite{Louge_Keast_2001,Lorenz_etal_1997}. The value of
the lower bound $\theta_{\textrm{min}}$ is not available in \cite{Walton1993}.
Velocity profiles as a function of distance from base at low angle
in \cite{Walton1993} show a seemingly linear shearing in the bulk
region (above the rolling layer - see Fig.~\ref{fig:Walton}) for
thin flows, which turn into constant-velocity crystallized plugs at
larger thickness.

At larger angles but in 2D \cite{Campbell_Brennen_1985}, the profiles
of the packing fraction, the velocity and its fluctuations, are of
the same type as these predicted by kinetic theories (type II in \cite{Ahn_etal_1992}),
with an inversion of the density profile and a higher ``granular
temperature'' at the base than at the surface. The state of art on
granular flows down flat frictional channels thus remain largely incomplete,
with limited numerical simulations not able to complement and detail
the inner details of the flows reported in experimental works.

\section{Simulation method}

\begin{figure}
\begin{centering}
\includegraphics[width=1\columnwidth]{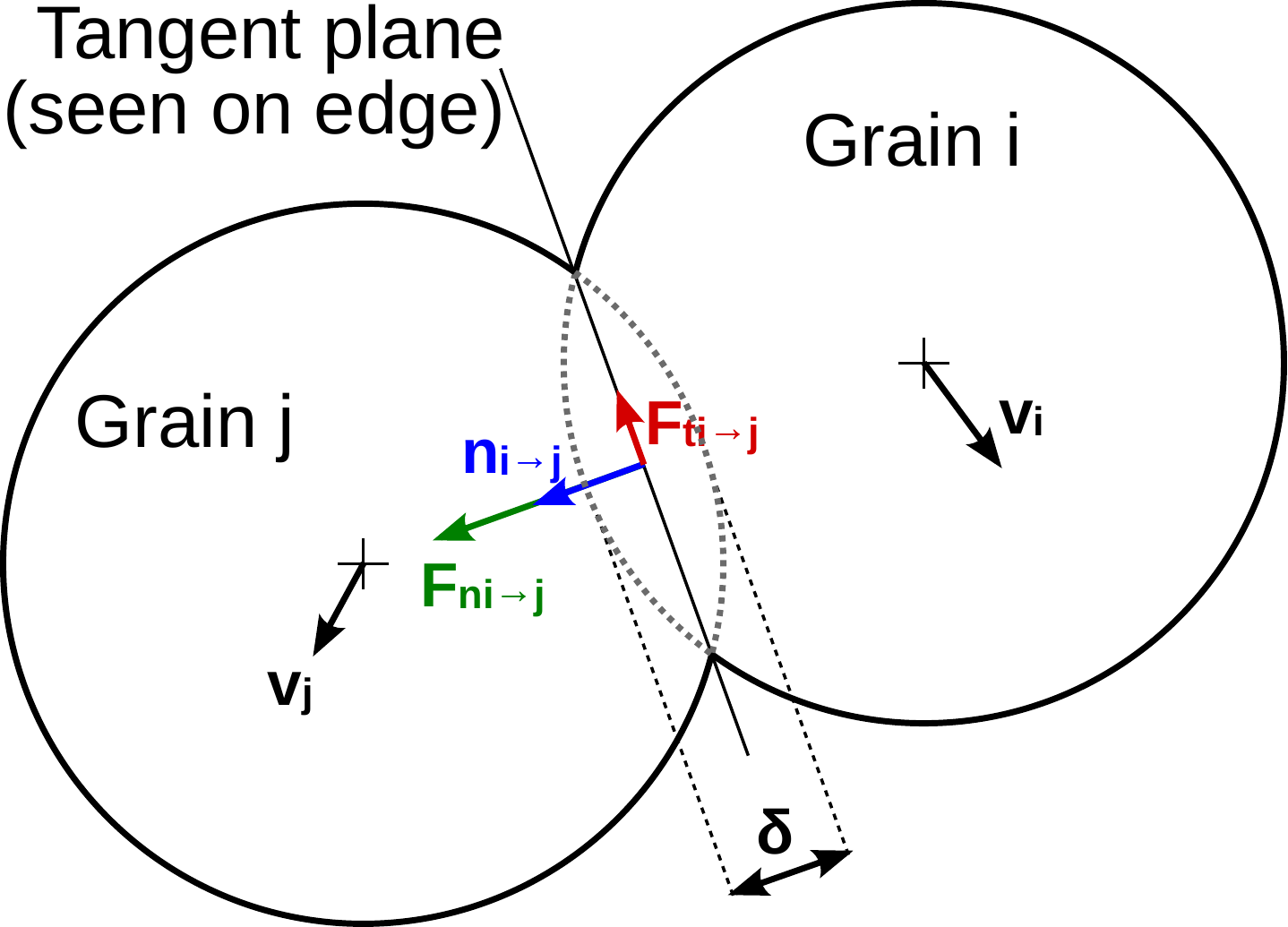} 
\par\end{centering}

\caption{\label{fig:OverlappingSpheres}(Color online). Overlapping spheres
contact model. Grain $i$ moves with a translation velocity $v_{i}$,
and similarly for $j$. The force exerted by grain $i$ on grain $j$
during contact (characterized by the normal vector $n^{i\rightarrow j}$)
is decomposed into a tangential component $F_{t}^{i\rightarrow j}$
and a normal component $F_{n}^{i\rightarrow j}$. Both depend on the
overlap $\delta$ according to a contact model detailed in the main
text.}
\end{figure}

We perform 3D numerical simulations of granular flows using molecular
dynamics (MD) that consist in integrating the equations of motion
over time. Each grain is represented by a sphere whose diameter is
drawn from a uniform distribution around the mean value $D$. The
grain model consists of a non-deformable sphere \cite{Cundall_Strack_1979}
of uniform material with density $\rho$. Deformations are taken into
account by the contact model, which links the normal force $\mathbf{F}_{n}$
acting on each grain to the overlap $\delta$ that occurs between
the non-deformed spheres when the grains centers are closer than their
diameters would allow (Fig. \ref{fig:OverlappingSpheres}). The linear
visco-elastic approach \cite{Luding2008} is used: $\mathbf{F}_{n}^{i\rightarrow j}=\left(k_{n}\delta+\gamma_{n}v_{n}\right)\mathbf{n}^{i\rightarrow j}$,
with $\mathbf{n}^{i\rightarrow j}$ the contact normal (unit vector
from sphere centers $i$ to $j$), $v_{n}=\left(\mathbf{v}_{i}-\mathbf{v}_{j}\right)\cdot\mathbf{n}^{i\rightarrow j}$
the normal component of the relative translational grain velocities,
$k_{n}$ a model spring stiffness and $\gamma_{n}$ a model viscosity
(i.e. giving a linear velocity-dependent force). A similar model is
applied in the tangential direction: $\mathbf{F}_{t}^{i\rightarrow j}=\left(k_{t}s+\gamma_{t}v_{t}\right)\mathbf{t}^{i\rightarrow j}$,
with $v_{t}\mathbf{t}^{i\rightarrow j}=\left(\mathbf{v}_{i}-\mathbf{v}_{j}\right)-v_{n}\mathbf{n}^{i\rightarrow j}$
the tangential component (i.e. with some direction within the tangential
plane) of the impact velocity, $k_{t}$ and $\gamma_{t}$ a model
spring stiffness and a model viscosity, and $s$ is a bounded version
$\left|s\right|\leq\left|\mathbf{F}_{t}\right|/k_{t}$ of the sliding
displacement $\int_{\tau_{0}}^{\tau}v_{t}d\tau$ in the tangential
plane since contact time $\tau_{0}$ \cite{Kruggel-Emden2008}. Coulomb
friction $\left|\mathbf{F}_{t}\right|\leq\mu\left|\mathbf{F}_{n}\right|$
is enforced on the tangential component, with a model coefficient
$\mu$. Below that threshold the value of $\mathbf{F}_{t}$ is given
by the above equations. The torque acting on a grain is computed as
$\mathbf{q}=-r\left(\mathbf{F}_{t}\times\mathbf{n}\right)$ with $r$
the grain radius. Both force and torque are used for integrating the
equation of motions $\sum\mathbf{F}=m\mathbf{a}$ and $\sum\mathbf{q}=I\dot{\boldsymbol{\omega}}$
with $m$ the mass of a grain, $\mathbf{a}$ its acceleration, $I$
its moment of inertia, and $\boldsymbol{\omega}$ its angular velocity
vector. Numerical integration is performed using the Velocity-Verlet
scheme. This whole approach is repeated for grain-wall interactions
with a different set of parameters $k_{n}^{gw},k_{t}^{gw},\gamma_{n}^{gw},\gamma_{t}^{gw},\mu^{gw}$.

Solid mechanics induces relations between these model parameters.
For a normal collision between two grains the damped harmonic oscillator
defined by the above interaction model leads to a contact duration
$\tau_{c}$ during which $\delta>0$ (half of the first pseudo-period).
Normal relative velocities before and after contact are then related
by a constant coefficient of normal restitution $e_{n}$ that sets
$\gamma_{n}$. Similarly the tangential spring/dashpot model defines
a coefficient of restitution $e_{t}$. Equating both duration times
leads to a relation $7k_{t}\left(\pi^{2}+(\ln\, e_{n})^{2}\right)=2k_{n}\left(\pi^{2}+(\ln\, e_{t})^{2}\right)$,
which corrects the $7k_{t}=2k_{n}$ relation from \cite{Silbert_etal_2001}
when $e_{n}\neq e_{t}$. Thanks to these relations the simulation
parameters can be concisely given in Table~\ref{tab:Simulation-parameters}.
\begin{table}
\begin{centering}
\begin{tabular*}{1\columnwidth}{@{\extracolsep{\fill}}|@{\extracolsep{\fill}}@{\extracolsep{\fill}}|l|l|}
\hline 
{\footnotesize Grain diameter}  & {\footnotesize $D=2.968$ (mm)}\tabularnewline
\hline 
{\footnotesize Grain mass (glass density $\rho_{g}$)}  & {\footnotesize $m=3.42\cdot10^{-5}$ (kg)}\tabularnewline
\hline 
{\footnotesize Gravity}  & {\footnotesize $g=9.81\textrm{ (m/s}^{2})$}\tabularnewline
\hline 
{\footnotesize Grain/grain normal restitution }  & {\footnotesize $e_{n}^{gg}=0.972$}\tabularnewline
\hline 
{\footnotesize Grain/grain tangential restitution}  & {\footnotesize $e_{t}^{gg}=0.25$}\tabularnewline
\hline 
{\footnotesize Grain/wall normal restitution}  & {\footnotesize $e_{n}^{gw}=0.8$}\tabularnewline
\hline 
{\footnotesize Grain/wall tangential restitution}  & {\footnotesize $e_{t}^{gw}=0.35$}\tabularnewline
\hline 
{\footnotesize Grain(glass)/grain friction}  & {\footnotesize $\mu^{gg}=0.33$}\tabularnewline
\hline 
{\footnotesize Grain(glass)/wall(aluminum)}  & {\footnotesize $\mu^{gw}=0.596$}\tabularnewline
\hline 
{\footnotesize Grain/grain spring stiffness}  & {\footnotesize $k_{n}^{gg}=2\cdot10^{5}\,(mg/D)$}\tabularnewline
\hline 
\multicolumn{2}{|c}{{\footnotesize Grain/wall $k_{n}^{gw}=k_{n}^{gg}$ (glass Young modulus
= aluminum)}}\tabularnewline
\hline 
{\footnotesize Integration time step} & {\footnotesize $dt=10^{-4}$ ($\sqrt{gD}$)}\tabularnewline
\hline 
\end{tabular*}
\par\end{centering}

\caption{\label{tab:Simulation-parameters}Simulation parameters matching the
experimental values from \cite{Louge_Keast_2001}.}
\end{table}

The correspondence of these parameters to physical values is subject
to a few simplifications. The most drastic one is the use of a single
model friction coefficient $\mu$ for all cases of static, kinetic
and collisional frictions. We had to use the static friction coefficient
instead of the other ones -- as usually done in MD simulations \cite{Borzsonyi_etal_2009,Silbert_etal_2001}
-- in order to reproduce experimental values. Hypothesis for this
model/experiment discrepancy given in the literature are the presence
of long lasting contacts \cite{Louge_Keast_2001} or the use of the
visco-elastic contact model itself \cite{Pournin_etal_2001}. Even
then, the static friction coefficient is known to be quite sensitive
to the surface properties and its determination is itself a topic
of debate. We used $\mu^{gg}=0.33$ as measured in our lab between
spent glass beads. Lorenz et al. \cite{Lorenz_etal_1997} had to erode
grains by circulating them in their experimental facilities for two
hours before their results became reproducible. We used the value
they give $\mu^{gw}=0.596$ for the grain/wall contacts in our simulations,
together with all their normal and tangential restitution measurements
(see Table \ref{tab:Simulation-parameters}). These restitution coefficients
could be refined for binary collisions using precise velocity-dependent
measures fitted by more complicated models \cite{Kruggel-Emden2008},
but this would not necessarily give better global results given the
multiple- and long-lasting- contacts \cite{Kruggel-Emden2008}, so
we stick to the experimental values given by \cite{Lorenz_etal_1997,Louge_Keast_2001}.
Similarly the use of a more complicated non-linear contact laws (\textit{e.g.}
Hertz) has been proposed but was found to be no better than a linear
model on a global scale \cite{Renzo_Maio_2004}. The value of the
spring stiffness shall however be related to material properties.
A link to the Young modulus and Poisson ratio is possible for Hertzian
contacts \cite{KruggelEmden2007}. For linear models we had to rely
on an ad-hoc approximation \cite{Richard_etal_2012} that leads to
$k_{n}=3\cdot10^{5}\textrm{N/m}=3.35\cdot10^{6}mg/D$ (comparable
to $k^{n}=2.10^{5}$N/m in \cite{Haynes_Walton_2000}). In any case
we checked that our results are not sensitive to the choice of $k_{n}$
provided it is given a sufficiently high value, so we then used the
more classical value $k_{n}=2.10^{5}\, mg/D$ for faster simulations
\cite{Silbert_etal_2001}.

\section{Flow Configuration}

\subsection{General setup}

\begin{figure}[t]
\begin{centering}
\includegraphics[width=1\columnwidth]{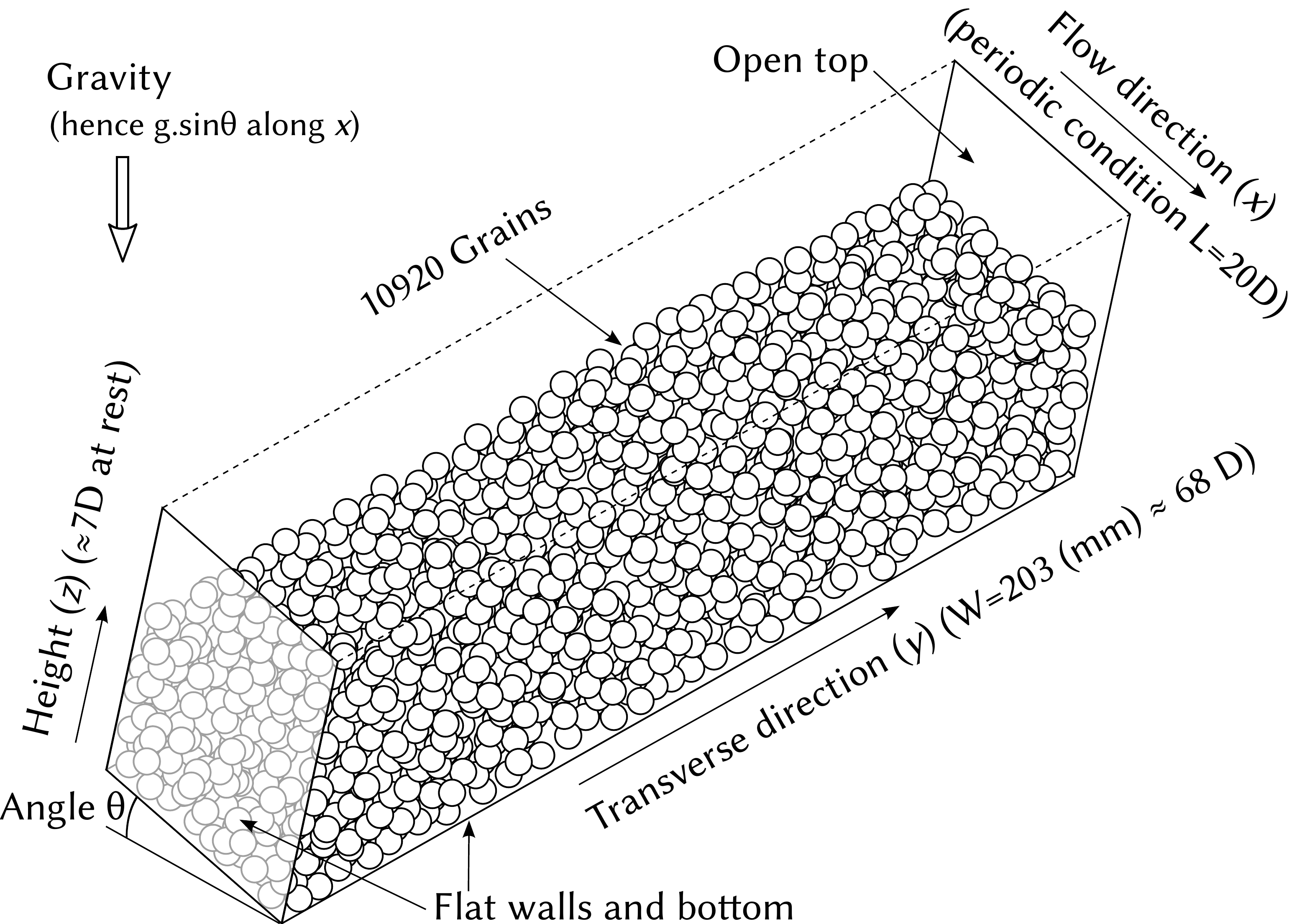} 
\par\end{centering}

\caption{\label{fig:Scenario-description}Sketch of the MD simulation in a
configuration corresponding to experiments \cite{Louge_Keast_2001}
where an inclined plane is bounded by flat side walls and base.}
\end{figure}

The simulation setup is designed to model the experimental setup of
Louge and Keast \cite{Louge_Keast_2001}, with minor adaptations (Fig.
\ref{fig:Scenario-description}). The calculational space is bounded
on the base and on the side walls by fixed flat frictional planes,
and it is free on the top surface (see Fig. \ref{fig:Scenario-description}).
PBC are applied in the flow ($x$) direction as we cannot simulate
the whole system with current computational facilities (we use a period
of $L=20D$, similar to \cite{Silbert_etal_2001}). Initial conditions
model the dropping of a loose assembly of agitated grains at a small
altitude. These low energy conditions are combined with a mass holdup
$\tilde{H}=4$ compatible with the experimental configurations for
all SFD flows in \cite{Louge_Keast_2001} (the mass holdup $\tilde{H}=\int_{0}^{+\infty}\frac{\nu(z)}{D}dz$
quantifies the amount of matter above a unit surface, with $\nu$
the volume fraction).

Preliminary simulations using PBC along $y$ and a small periodic
cell size were first performed and were able to recover the results
of Walton~\cite{Walton1993}. However, the range of angles of inclination
for which steady and fully developed flows are reached does not match
the experimental results of Louge and Keast~\cite{Louge_Keast_2001}.
For example, in simulations we obtain $\theta_{\textrm{min}}\approx6-7$\textdegree{},
which is much lower than the experimental value ($15.5$\textdegree{}).
Neither a modification of the material parameters such as the friction
and the restitution coefficients, the polydispersity of grains, nor
the introduction of other sources of dissipation like rolling friction
\cite{Luding2008} gives values of $\theta_{\textrm{min}}$ close
the experimental one. On the contrary, the introduction of side walls
separated by a gap $W$ with identical material properties as the
base is able to increase $\theta_{\textrm{min}}$ to $\theta_{\textrm{min}}=14^{\circ}$,
reasonably close to the experimental value ($15.5^{\circ}$). Note
also that, the use of the low polydispersity value given in \cite{Louge_Keast_2001}
($\pm0.7\%$ of the average grain diameter) leads to crystallized
blocks in the flow. Due to the periodicity in $x$ these blocks tend
to persist for a long time. Experimentally the grains are re-circulated
and this corresponds to averaging over multiple realizations making
the presence of blocks inappropriate. One way to get rid of these
artifacts was to increase the polydispersity, up to 10\% for all the
results presented below.

\subsection{The flow regimes}

\begin{figure}
\begin{centering}
\includegraphics[width=1\columnwidth]{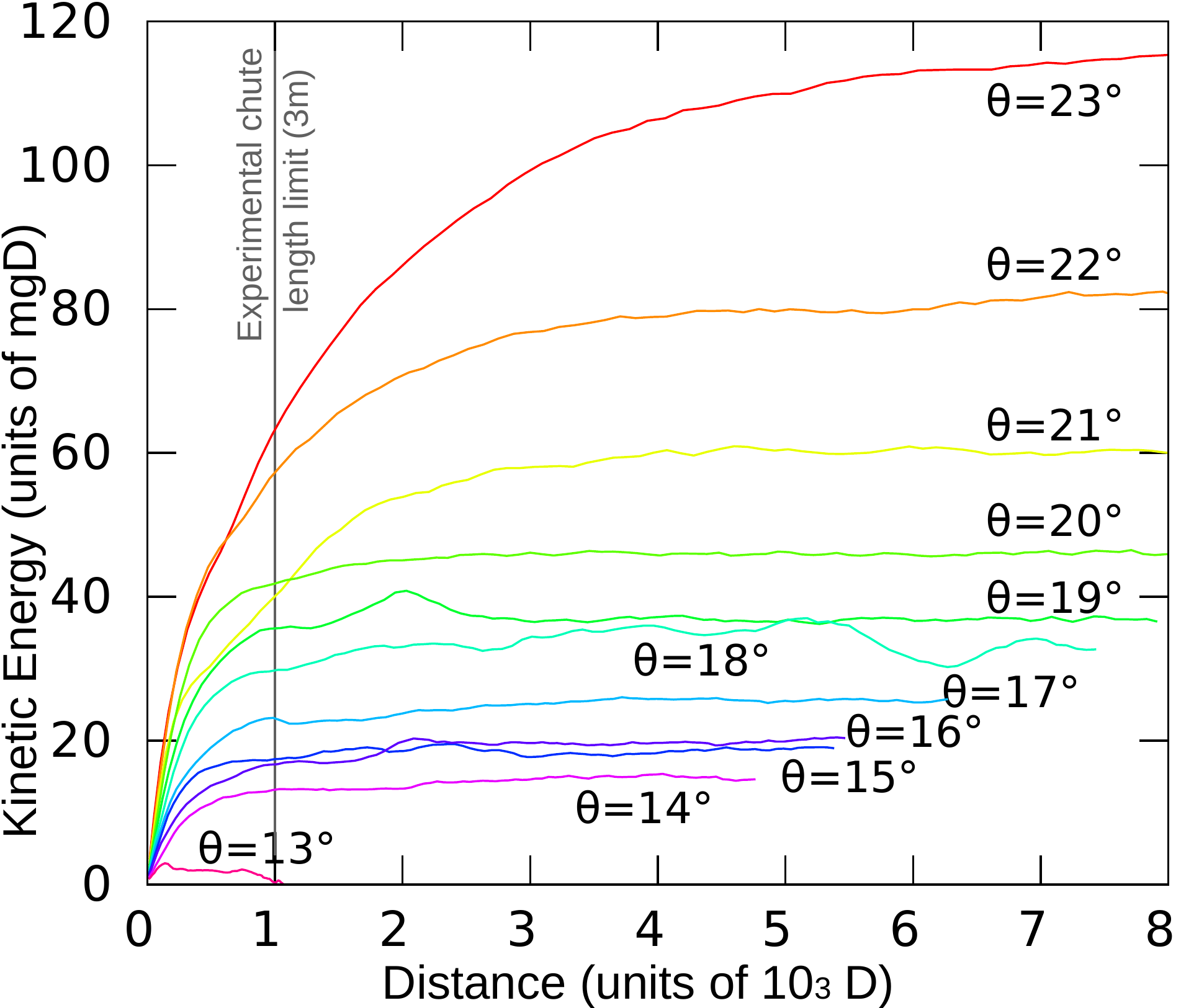} 
\par\end{centering}

\caption{\label{fig:KineticEnergy}(Color online). Translation kinetic energy
vs distance traveled by the grains, showing regimes that appear accelerated
within the chute length experimental limit, but reaching steady states
at larger distances.}
\end{figure}

We ran simulations for a range of angles from 13\textdegree{} to 23\textdegree{}
containing the range of experimentally observed SFD flows \cite{Louge_Keast_2001}.
Visual investigation of the simulations shows that the main bulk of
the flows rests on top of a basal layer of grains, for which there
is a combination of long-lasting rolling contacts with the flat base
that are interupted by short rebounds. Fig.~\ref{fig:KineticEnergy}
presents the evolution of the kinetic energies of the flows over the
average distance traveled by the grains. Louge and Keast \cite{Louge_Keast_2001}
experimentally observed a range of angles for steady states from 15.5\textdegree{}
to 20\textdegree{}, established over distances less than 3m. This
matches our finding (see the vertical line in Fig. \ref{fig:KineticEnergy})
that flows at higher angles would indeed appear accelerated within
the experimental limits. The flows stop below $\theta_{\textrm{min}}=14$\textdegree{}
which reasonably matches the experimental value (15.5\textdegree{})
as there are extra factors not taken into account in the simulation,
like the abundant inter-particle dust that was reported experimentally
\cite{Louge_Keast_2001,Lorenz_etal_1997}, which might then block
the flow at low velocities. The effect of the side walls on $\theta_{\textrm{min}}$
mentioned by Louge and Keast~\cite{Louge_Keast_2001} cannot be attributed
only to additional friction, which is negligible given the shallow
flow height and the large distance between the walls as aforementioned
and noted in \cite{Louge_Keast_2001}. Some unknown mechanism is thus
at work, which will be the topic of further studies.

Simulations with sidewalls lead to SFD flows except, maybe, for $\theta\approx18\text{\textdegree}$
($\pm1\text{\textdegree}$) where relatively large fluctuations are
visible on the kinetic energy in Fig. \ref{fig:KineticEnergy}. These
fluctuations, which persist over time (we checked it up to a distance
of $\approx13500D$), were also reported experimentally in \cite{Louge_Keast_2001},
although it is not easy to determine whether these match our simulations
results: the periodic size we use in $x$ is a fraction of the oscillation
wavelength observed experimentally. Hanes and Walton \cite{Haynes_Walton_2000},
who use a bumpy base for their experiments, also report a phase diagram
with an oscillating regime delimited by fuzzy boundaries, at the junction
of two SFD regimes, with the same PBC numerical interpretation difficulty.
These oscillations take place at the transition between two SFD regimes
which have very different behaviors. For $\theta\in\left[14\text{\textdegree},17\text{\textdegree}\right]$
flows are unidirectional and grains from ordered layers. These are
visible in Fig. \ref{fig:PackingFractionZ} as regions of higher packing
fraction, as well as in Fig. \ref{fig:packfracYZ}a. The free surface
of the flow (Fig. \ref{fig:packfracYZ}a) is convex, higher in the
center than on the borders. For $\theta$ larger than 19\textdegree{},
secondary flows develop (Fig. \ref{fig:Flow_and_T}), breaking the
layer structure (Fig. \ref{fig:PackingFractionZ}), except for the
basal layer of rolling grains (Fig. \ref{fig:packfracYZ}b). The free
surface is concave (Fig. \ref{fig:packfracYZ}b). The flow height
is enlarged, with a correspondingly lower average density, but the
flow remains shallow (height $\approx10D$ for a width of $68D$).
The rolls are thus quite flat. Secondary flows have only been observed
in experiments within a bumpy channel, \textit{e.g.} \cite{Savage1979}
(using $D=0.5$mm beads) and \cite{Borzsonyi_etal_2009} ($D=0.4$mm).
The stationary state in \cite{Savage1979} was however reached at
a distance compatible with our results, scaled by the difference in
$D$: less than 3m at $\theta=23.6$\textdegree{}.

In order to quantify the effect of the geometry of the base, we have
carry out the same simulations with a bumpy base consisting of fixed
grains and otherwise the same parameters (including flat frictional
lateral walls). In these conditions the grains flow only above 22\textdegree{},
with an average kinetic energy of $\approx1mgD$. Compared to the
kinetic energy $\approx100mgD$ in Fig. \ref{fig:KineticEnergy} at
the same angles, and given the much lower $\theta_{\textrm{min}}$
bound in the flat case, we immediately see that no direct comparison
is possible between the flat frictional base and the bumpy one: the
basal layer of rolling grains significantly reduces the dissipation.
Experimentally \cite{Savage1979} also noted a much increased flow
velocity on flat surfaces. For $\theta\geq22\text{\textdegree}$ fixing
grains on the base prevents the rolls but induce a large internal
agitation instead. This is consistent with the results of Börzsönyi
\textit{et al.}~\cite{Borzsonyi_etal_2009} which report rolls only
for thicker flows compared to our shallow flow configuration. Therefore,
the convection regime is accessible for smaller systems with a flat
frictional bottom than with a bumpy one.

\subsection{Influence of the parameters and of the initial conditions}

To study the generality of the above reported results we carry out
an extensive study of the effect of the model parameters and of the
initial conditions which is summarized below.\\
 The transitions between the regimes and their characteristics depend
slightly on the model parameters (\textit{e.g.} friction coefficients,
polydispersity of the grains) but the general features of the flows
seem robust. For instance, additional runs with $\mu_{gg}=0.4$ instead
of $0.33$ shifted the start and end of the unidirectional flows up
by 1\textdegree{}. The rolls are robust to polydispersity (tested
with $D\pm20\%$). They also appear when $\mu_{gg}=\mu_{gw}$ (customary
setting in numerical works \cite{Silbert_etal_2001}), provided both
are greater than about 0.54.

The initial conditions we use consist of dropping a loose assembly
of grains at $z=2D$ with a low initial velocity and some jitter,
which we designed to be approximately what the grains would have experimentally
when leaving the open gate in \cite{Louge_Keast_2001}. The corresponding
initial energy is much lower than that reached in steady state (see
Fig. \ref{fig:KineticEnergy}). We checked the final SFD states are
robust to variations of the initial conditions provided these induce
an initial energy smaller than the energy of the final SFD state.
However we have not studied the use of larger energies, as we suspect
there may be hysteresis effects on the final energy levels. Literature
for the bumpy base case also reports \cite{Taberlet2007} that specific
regimes exist with high initial energy conditions, so presumably this
might be a possibility for the flat frictional base case as well.

We varied the mass holdup so as to match the range of shallow flow
configurations in \cite{Louge_Keast_2001}: low $\tilde{H}=1$ induces
an early transition to a dilute phase without secondary flows. Convection
rolls appear between $\tilde{H}=3$ and $\tilde{H}=4$. They persist
even for much higher mass holdups (tested up to $\tilde{H}=20$).

The existence of a minimal angle $\theta_{\min}$ for SFD flows requires
the presence of enough layers of grains, as the rolling basal layer
may accelerate indefinitely with insufficient frustrations on the
grain rotations (the limit case being a single grain rolling on an
flat inclined plane). The investigation of the parameters and initial
conditions mentioned above shows that the properties of the flows
are robust. Therefore, the simulations reported here are qualitatively
representative of many others obtained with different conditions.

\section{Steady states}

This section analyzes the main features of the two families of SFD
states : unidirectional and layered flows, and flows with granular
convection. All the figures presented below report time averaged quantities
(\textit{e.g.} velocity, packing fraction) computed over 500 $\sqrt{D/g}$
time units in steady state, as well as over the periodic cell in the
flow direction. These results are stable with respect to the particular
random seed we use between different runs.

\subsection{\label{sub:Packing-fraction}Packing fraction}

\begin{figure}
\begin{centering}
\includegraphics[width=1\columnwidth]{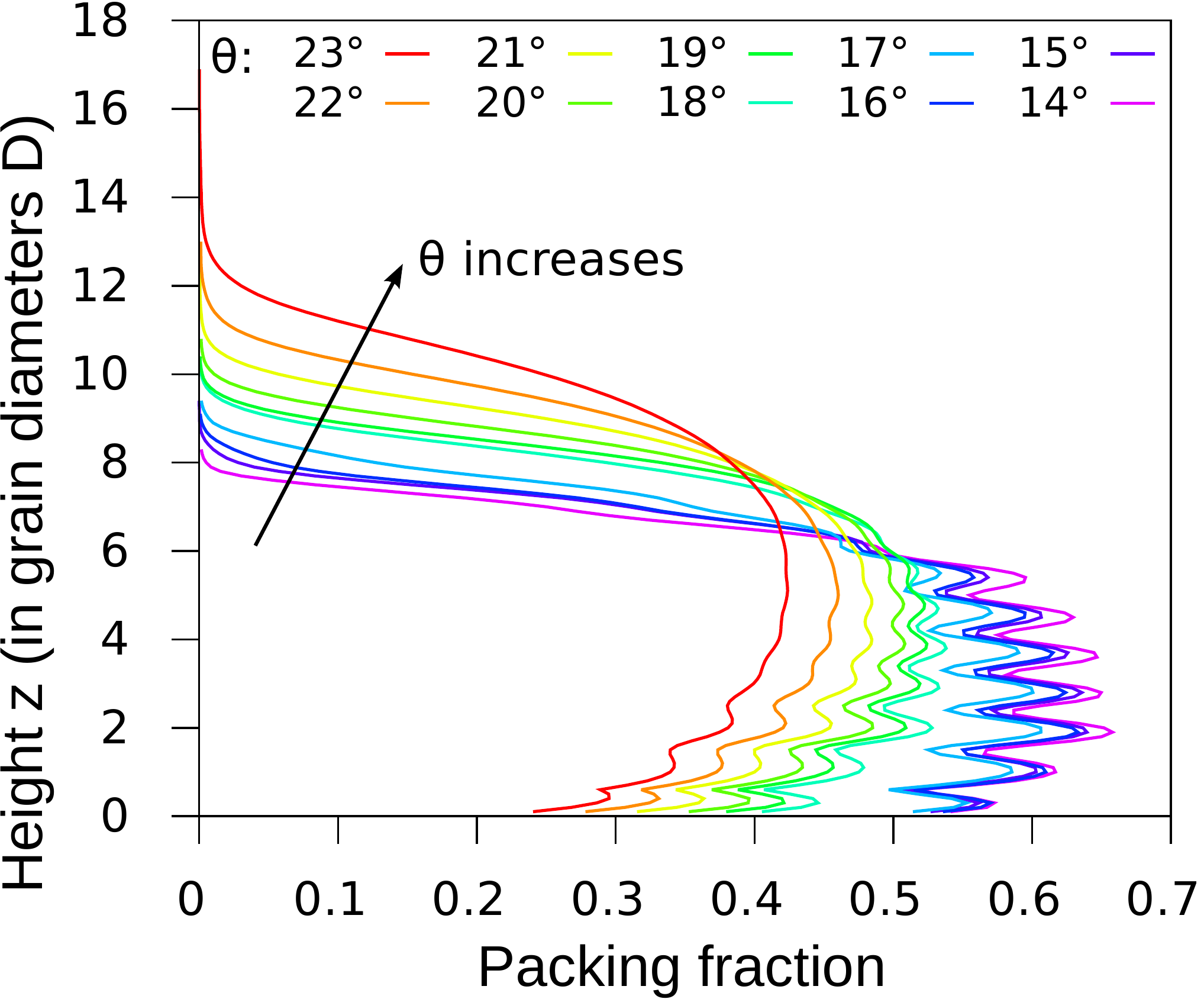} 
\par\end{centering}

\caption{\label{fig:PackingFractionZ}(Color online). Height-averaged packing
fraction (smoothed over $\pm0.5D$ in $z$) for each angle of inclination.
The layering is clearly apparent at low angles, and disappears in
the presence of the secondary flows.}
\end{figure}

\begin{figure*}
\begin{centering}
\includegraphics[width=1\textwidth]{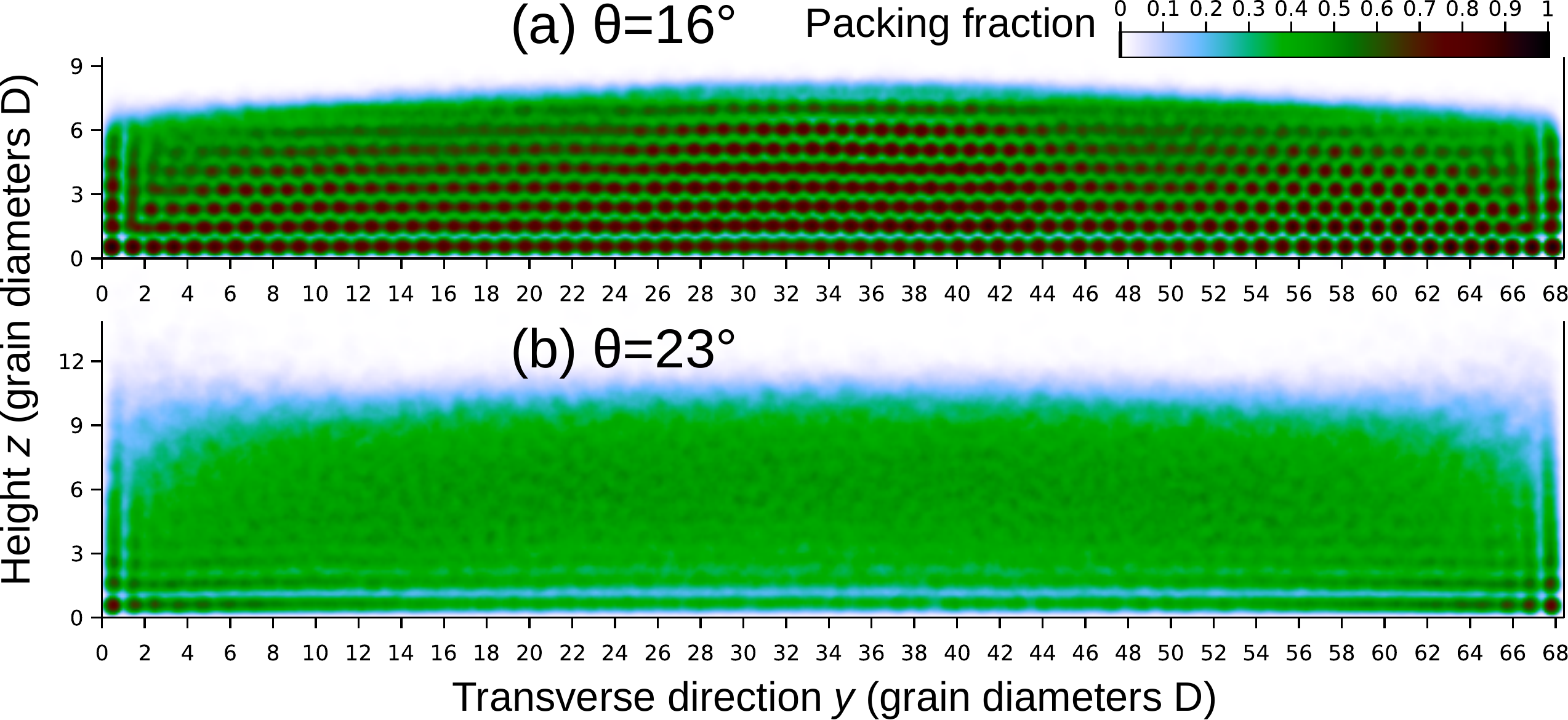} 
\par\end{centering}

\caption{\label{fig:packfracYZ}(Color online). Packing fraction map, averaged
over $x$. The layered structure at low angles disappears when granular
convection takes place.}
\end{figure*}

Values of the packing fraction at the base of the flow (Fig. \ref{fig:PackingFractionZ})
are compatible with the experiments (see for example Fig. 6 of \cite{Louge_Keast_2001},
$\theta=16\text{\textdegree},18\text{\textdegree},20\text{\textdegree}$
and $\tilde{H}=4$). For the unidirectional flows Fig. \ref{fig:PackingFractionZ}
shows that the average volume fraction $\nu\approx0.59$ does not
vary much in $z$ above the basal layer, with clear variations around
that average at each structured layer. The structuration in layers
of constant packing fraction (above the basal layer) was observed
by the previous numerical study \cite{Walton1993} with PBC along
$y$. We confirm that structure persists with a polydispersity of
10\%, that it was not an artifact of the use of single-sized spheres
in \cite{Walton1993}. Another difference with \cite{Walton1993}
is the presence of 2 lateral layers (Fig. \ref{fig:packfracYZ}) at
the wall boundaries, inducing some nearby structuration.

In the convective regime we observe an inverted density profile (Fig.
\ref{fig:PackingFractionZ}), similar to these reported in the literature
for bumpy boundaries \cite{Borzsonyi_etal_2009,Forterre_Pouliquen_2002},
while the layer in contact with the flat boundaries remains clearly
separated (Fig. \ref{fig:PackingFractionZ} and \ref{fig:packfracYZ}b).

\subsection{Pressure and effective friction}

\begin{figure}
\begin{centering}
\includegraphics[width=1\columnwidth]{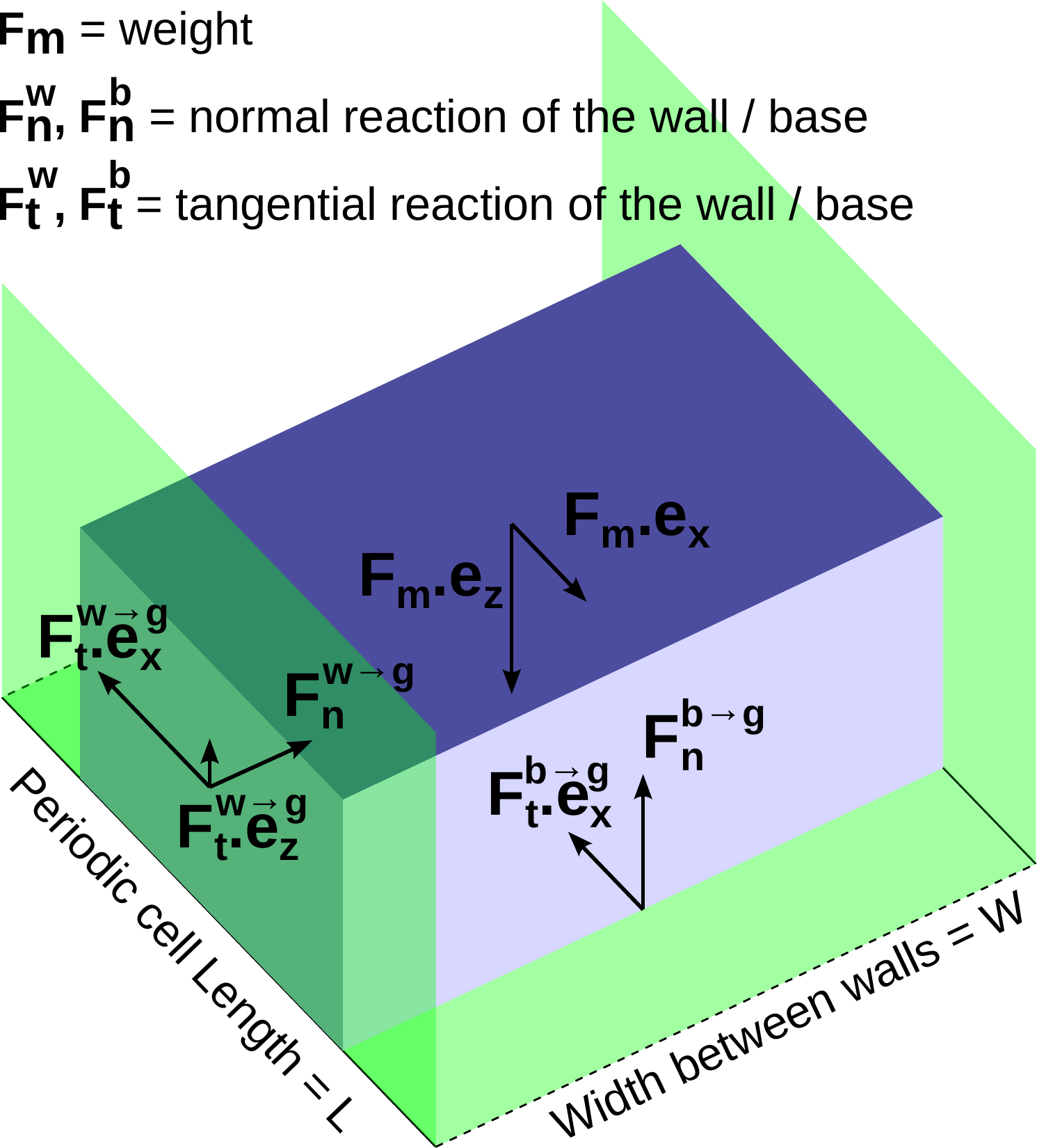} 
\par\end{centering}

\caption{\label{fig:Forces-budget}(Color online). External forces balance
on the grain flow.}
\end{figure}

\begin{figure}
\begin{centering}
\includegraphics[width=1\columnwidth]{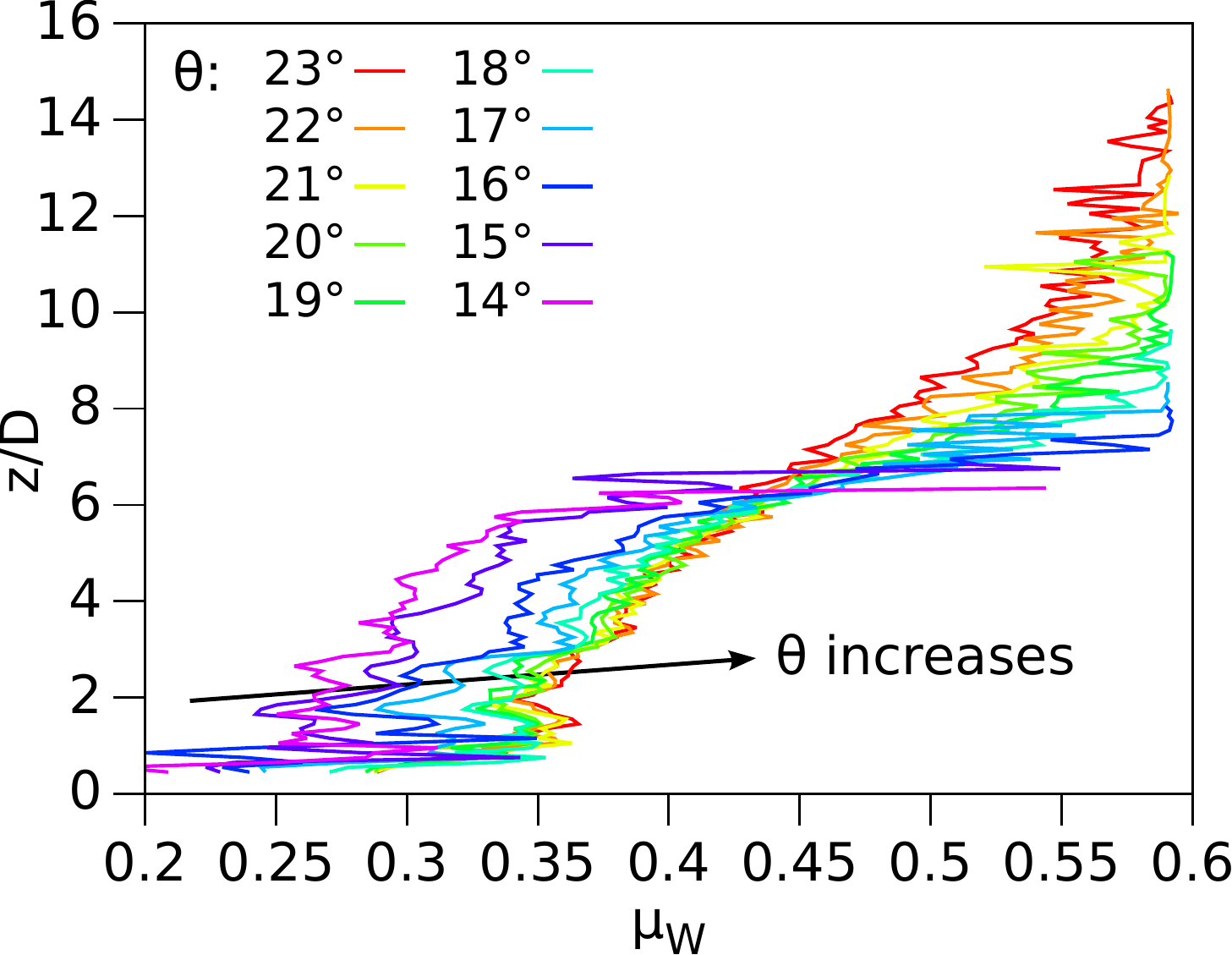} 
\par\end{centering}

\caption{\label{fig:muW}(Color online). Friction coefficient profile on the
lateral walls for several value of the inclination angle $\theta$.}
\end{figure}

We did not implement a full calculation of the stress tensor, but
stresses on the boundaries are easily deduced from the forces exerted
by the grains during each contact, averaged over a $500\sqrt{D/g}$
time window. Let us denote by $\mathbf{\mathbf{f}^{\mathit{g\rightarrow w}}\mathrm{(\mathit{z})}}$
the stress vector exerted on one wall, at height $z$, by the grains.
The force exerted by one wall on the grain flow is simply: $\mathbf{F}^{\mathit{w\rightarrow g}}=-\int_{x=0}^{L}\int_{z=0}^{\infty}\mathbf{\mathbf{f}^{\mathit{g\rightarrow w}}\mathrm{(\mathit{z})}}dxdz=L\int_{z=0}^{\infty}\mathbf{\mathbf{f}^{\mathit{w\rightarrow g}}\mathrm{(\mathit{z})}}dz$.
Using the subscript $n$ and $t$ to distinguish the normal and the
tangential components, the local effective friction coefficient on
the wall is then computed as $\mu_{W}(z)=\left\Vert \mathbf{f}_{t}(z)\right\Vert /\left\Vert \mathbf{f}_{n}(z)\right\Vert $.
The global effective friction coefficient is computed as $\widehat{\mu_{W}}=\left\Vert \mathbf{F}_{t}\right\Vert /\left\Vert \mathbf{F}_{n}\right\Vert $.
The profiles in $z$ and the values of the effective friction on the
walls are shown in Fig. \ref{fig:muW} and Table \ref{tab:=00003D00003D0003BC_W and R}.
The observed friction weakens with depth and is similar to that reported
in \cite{Richard2008}.

\begin{table}
\begin{centering}
\begin{tabular}{|c|c|c|c|c|c|c|c|c|c|c|}
\hline 
{\small $\theta$}  & {\small 14\textdegree{}}  & {\small 15\textdegree{}}  & {\small 16\textdegree{}}  & {\small 17\textdegree{}}  & {\small 18\textdegree{}}  & {\small 19\textdegree{}}  & {\small 20\textdegree{}}  & {\small 21\textdegree{}}  & {\small 22\textdegree{}}  & {\small 23\textdegree{}}\tabularnewline
\hline 
{\small $\widehat{\mu_{W}}$}  & {\small 0.27}  & {\small 0.28}  & {\small 0.31}  & {\small 0.32}  & {\small 0.35}  & {\small 0.36}  & {\small 0.36}  & {\small 0.37}  & {\small 0.38}  & {\small 0.38}\tabularnewline
\hline 
\end{tabular}
\par\end{centering}

\caption{\label{tab:=00003D00003D0003BC_W and R}Global effective friction
coefficient on the wall (to compare with the microscopic $\mu_{gw}=0.596$).}
\end{table}

\begin{figure}
\begin{centering}
\includegraphics[width=1\columnwidth]{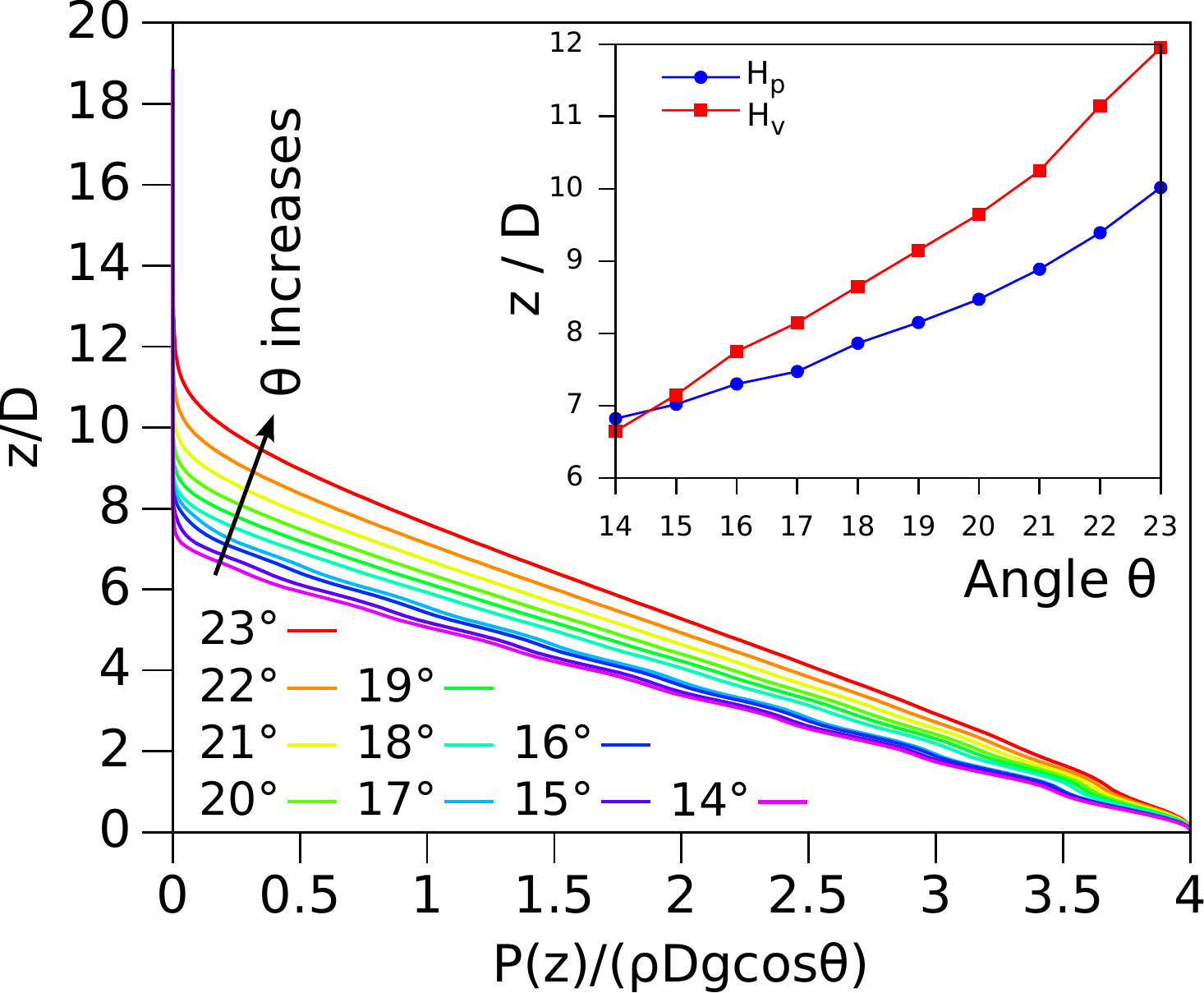} 
\par\end{centering}

\caption{\label{fig:P_Hp}(Color online). Main diagram: Vertical profiles of
the averaged pressure for the hydrostatic approximation in the center
part of the curves. Inset: The effective heights of the flows, with
respect to the hydrostatic approximation $H_{p}$ and maximal flow
velocity $H_{v}$. See the main text.}
\end{figure}

Let us consider a slab at the top of the flow (from $z$ and above)
as a continuum. The balance of external forces along $z$ implies:
$\left(\int_{x=0}^{L}\int_{y=0}^{W}\int_{z}^{\infty}\nu(x,y,z')dxdydz\right)\rho g\cos\theta=P(z)LW+2L\int_{z}^{\infty}\mathbf{f}_{t}^{w\rightarrow g}(z').\mathbf{e_{z}}dz'$
with $P(z)$ the average pressure on an $xy$ plane section computed
at height $z$. The contribution of the walls to this balance is always
very small compared to the other terms (of the order of one per cent).
The vertical profiles of $P(z)$ are shown in Fig. \ref{fig:P_Hp}
for each angle $\theta$, normalized so the value at the base equates
the mass holdup $\tilde{H}$ we use. They are linear over the most
part, except for the basal rolling layer and the very diluted top
consisting of a few grains in ballistic motion. Thus, over the main
bulk of the flow, the approximation $P(z)\approx\left(H_{p}-z\right)\overline{\nu}\rho_{g}g\cos\theta$
is excellent (with $\bar{\nu}$ the average packing fraction on the
bulk). The corresponding effective flow heights $H_{p}$ are shown
in Fig. \ref{fig:P_Hp} (inset). They confirm the general dilation
of the flow with the angle $\theta$, matching the general packing
fraction decrease of Fig. \ref{fig:PackingFractionZ}. The total frictional
influence of the walls on the flow can also be quantified. The ratio
between the weight of the grains and the friction force on the walls:
$\widehat{\mu_{W}}\: H_{p}/W$ \cite{Taberlet2003,Taberlet_etal_2005}
is at most $0.06$ for $\theta=23\text{\textdegree}$. This justifies
the arguments developed in \cite{Louge_Keast_2001} and used here
for neglecting the friction on the walls, for a shallow flow with
a large width.

\begin{figure}
\begin{centering}
\includegraphics[width=1\columnwidth]{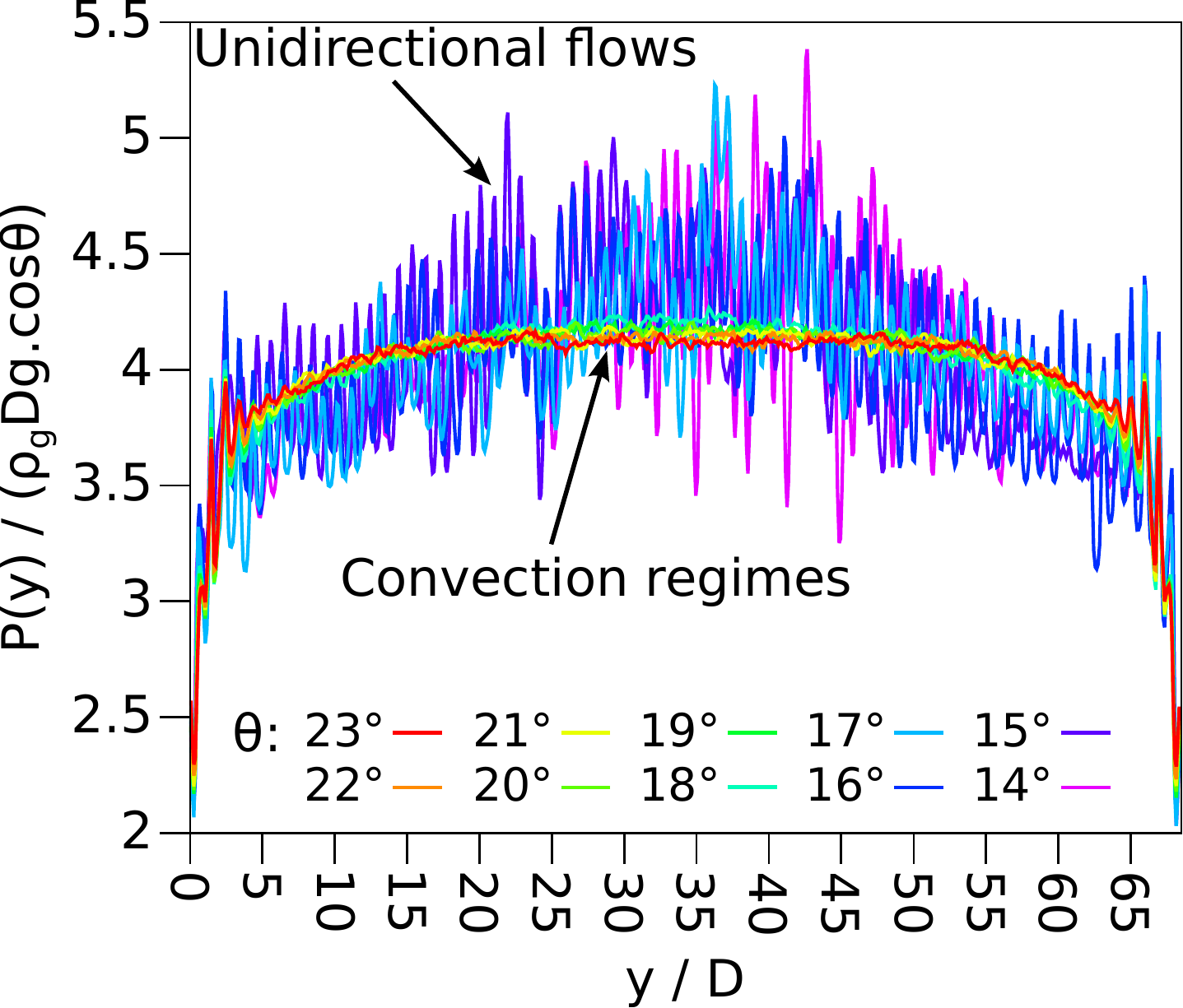} 
\par\end{centering}

\caption{\label{fig:Pbase}(Color online). Normalized pressure computed from
each contact with the base, smoothed over $\pm1D$ in $y$}
\end{figure}

The pressure at any position on the flat base $P_{B}(y)=\mathbf{f}_{n}^{g\rightarrow b}(y).\mathbf{e_{z}}$
is shown in Fig. \ref{fig:Pbase}, normalized so the average value
is the mass holdup $\tilde{H}=4$. In the unidirectional regimes grains
do not deviate much in $y$ from their trajectories along $x$, leading
to large pressure fluctuations on the base. In the convective regime,
the grain circulation due to the secondary flows smooth out these
differences. In each case pressure is maximal at the center of the
flow, and decreases in the lateral parts near the walls.

\subsection{Mean velocity and fluctuations}

\begin{figure}
\begin{centering}
\includegraphics[width=1\columnwidth]{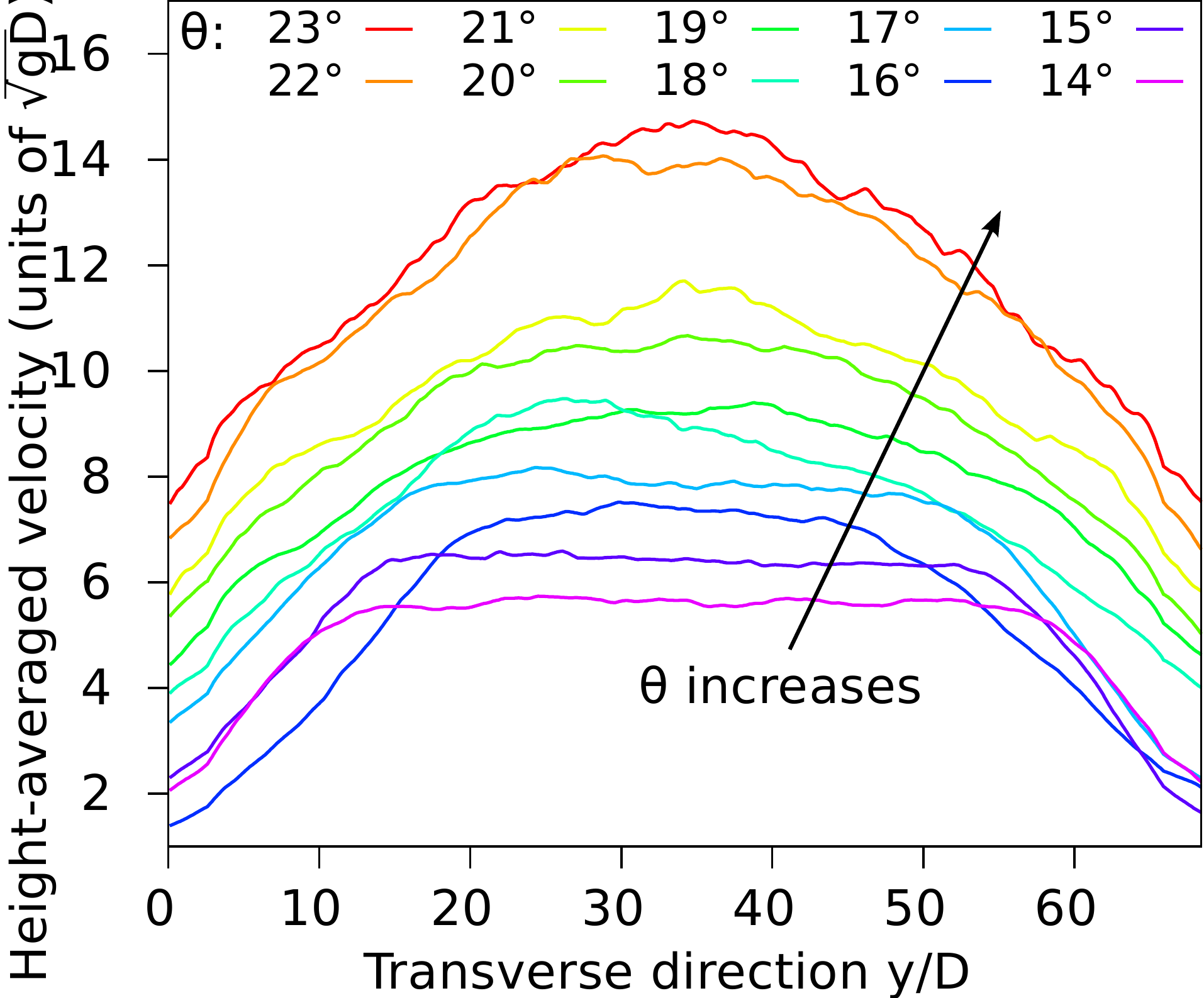} 
\par\end{centering}

\caption{\label{fig:vprofileY}(Color online). Transverse velocity profile,
smoothed over $\pm2.5D$ in $Y$, averaged over $z$.}
\end{figure}

\begin{figure}
\begin{centering}
\includegraphics[width=1\columnwidth]{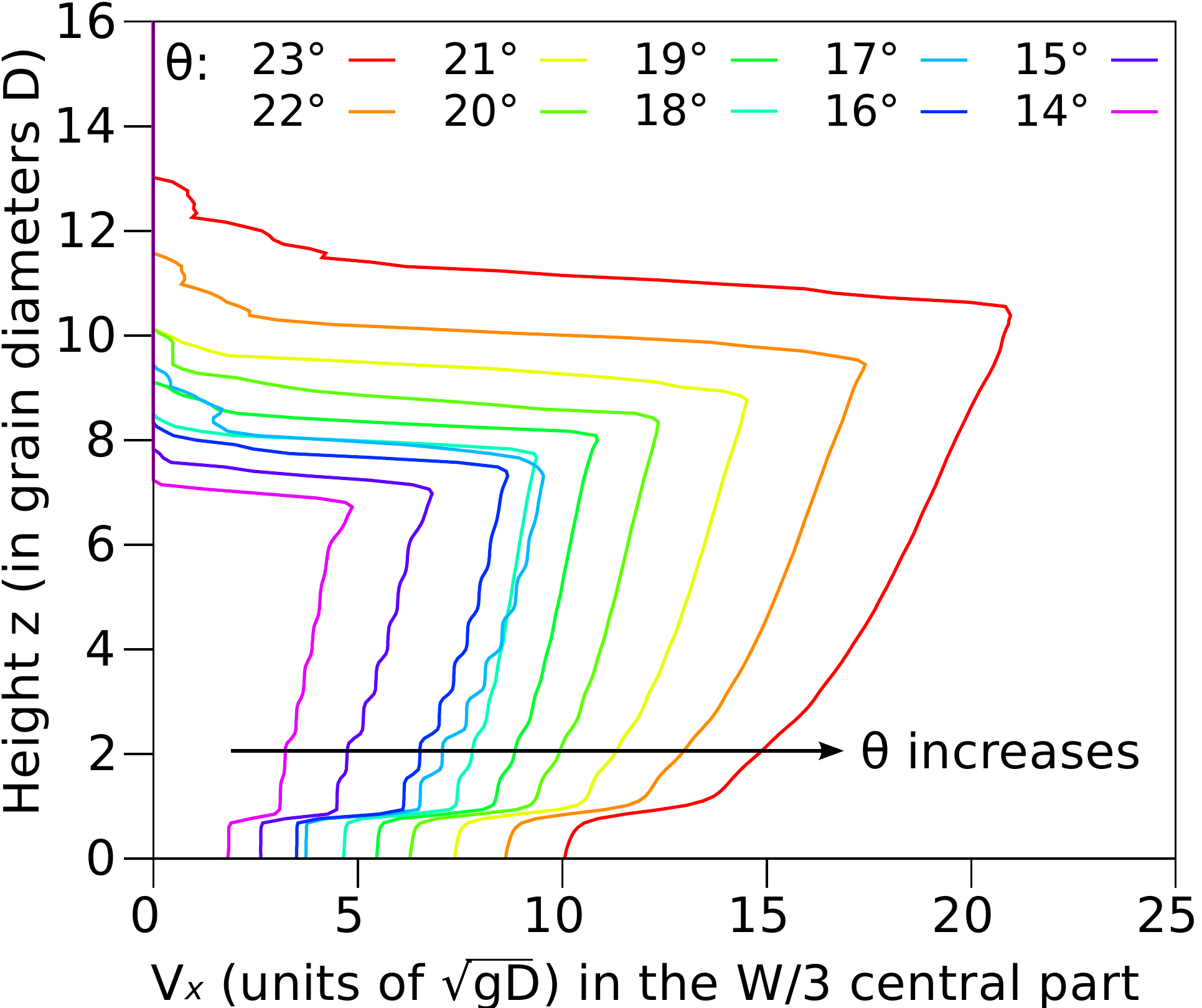} 
\par\end{centering}

\caption{\label{fig:vprofileZ}(Color online). Height profile of the velocity
along $x$ averaged over the central part of the flow.}
\end{figure}

A continuous mean velocity field $\overline{\mathbf{v}}$ is computed
using the definition by Serero et al. \cite{Serero_etal_2008} for
polydisperse systems, and averaged over time: $\overline{\mathbf{v}}\left(\mathbf{x}\right)=\left\langle \sum_{i=1}^{N}\mathbf{v}_{i}k_{i}\left(\mathbf{x}_{i}-\mathbf{x}\right)\right\rangle _{\tau}/\left\langle \sum_{j=1}^{N}k_{j}\left(\mathbf{x}_{j}-\mathbf{x}\right)\right\rangle _{\tau}$,
where $\mathbf{x}=(x,y,z)$ is the 3D position at which to compute
the average velocity, $\mathbf{x}_{i}$ is the position of the center
of grain $i$, and $k_{i}$ is a kernel that distributes the mass
$m_{i}$ of grain $i$ over space. We use the uniform density kernel
$k_{i}(\mathbf{x}_{i}-\mathbf{x})=\rho$ when $\left\Vert \mathbf{x}_{i}-\mathbf{x}\right\Vert <r_{i}$
with $r_{i}$ the radius of grain $i$, and $0$ elsewhere. We then
define a ``granular temperature'' \cite{Serero_etal_2008} from
the velocity fluctuations: $T=\frac{1}{2}\left(\overline{\left\Vert \mathbf{v}\right\Vert ^{2}}-\left\Vert \overline{\mathbf{v}}\right\Vert ^{2}\right)$
where the overline denotes the above weighted averaging.

For the unidirectional flows Fig. \ref{fig:vprofileY} reproduces
the experimental mean velocity transverse profile (Fig.~\ref{fig:Louge_Keast_2001_reprint}),
where the shearing layer induced by the walls extends to about 1/3
within the flow. The shape of the velocity profile in the unidirectional
regime, considering the average packing fraction is constant in $z$
(Fig. \ref{fig:PackingFractionZ}), is comparable to the experimental
measures in Fig. 3 (4) of \cite{Johnson_etal_1990} at similar angles.
Within the central part we obtain velocity profiles (Fig. \ref{fig:vprofileZ})
similar to these given by PBC \cite{Walton1993}. The velocity reaches
a maximal value at height $H_{v}$ then decreases rapidly in the sparse
ballistic layer of grains. The values of $H_{v}$ can be compared
to $H_{p}$ in the inset of Fig. \ref{fig:P_Hp}. The 3D velocity
profile of the flow can be inferred from Figs. \ref{fig:vprofileY}
and \ref{fig:vprofileZ} as a faster region in the center part, sheared
vertically, on top of a basal rolling layer of grains. The transverse
velocity profile (Fig. \ref{fig:vprofileY}) is sheared through the
whole width in the presence of secondary rolls. These convey grains
up to the center of the flow, as seen in Fig. \ref{fig:Flow_and_T}.

\begin{figure}
\begin{centering}
\includegraphics[width=1\columnwidth]{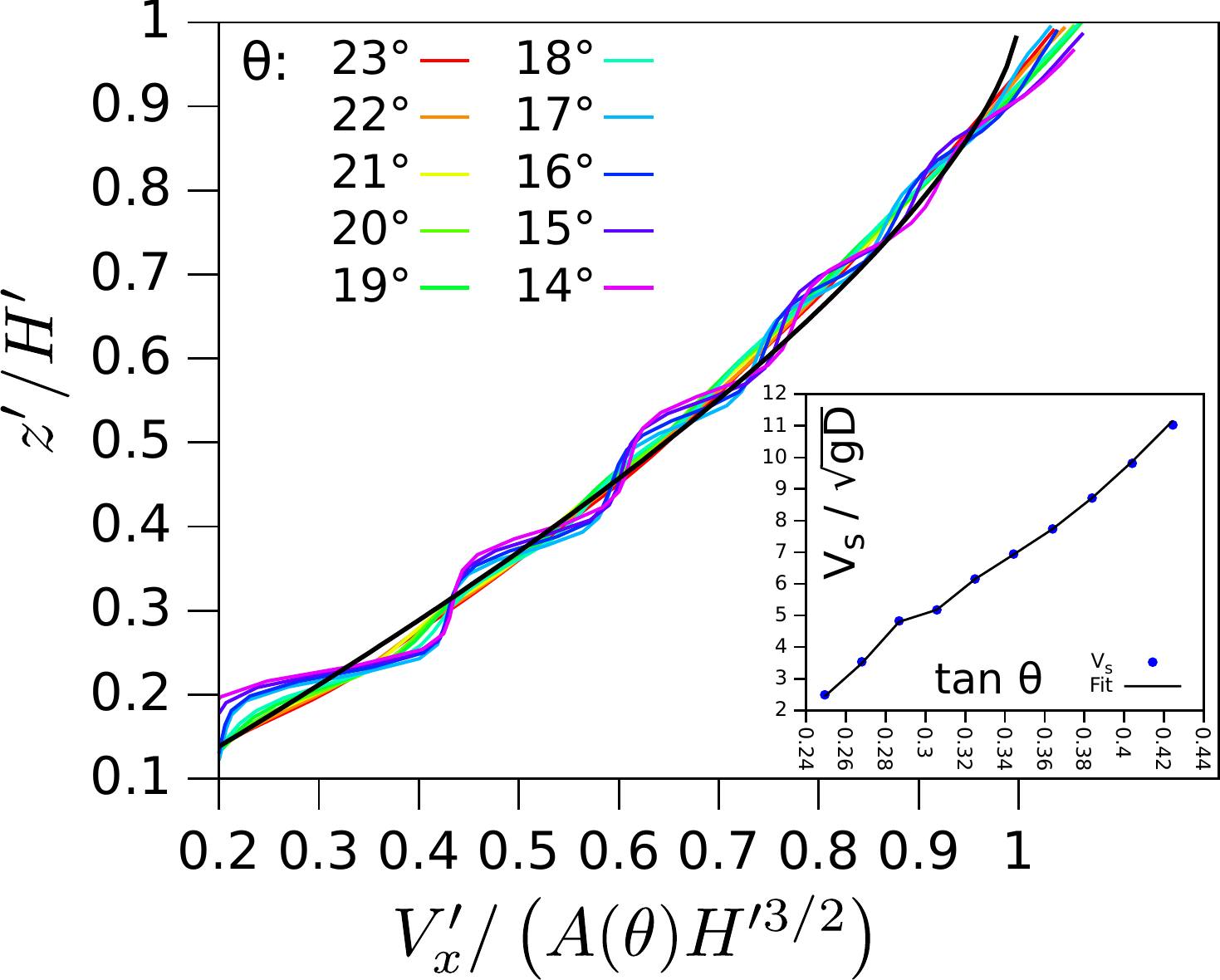} 
\par\end{centering}

\caption{\label{fig:Bagnold_collapse}(Color online). Main plot: Bagnold velocity
profile collapse. Inset: effective ``sliding'' velocities of the
main bulk of the flow with respect to the basal layer of rolling grains.}
\end{figure}

The basal layer of rolling and bumping grains can be interpreted as
an effective base for the main bulk of the flow on top of it. A sliding
velocity $V_{s}$ can be defined as the velocity in the direction
of flow at $z_{0}=1.5D$, corresponding to the mean velocity of the
grains in the second layer, just above the basal grains. Now, let
$V_{x}'=\left(V_{x}-V_{s}\right)/\sqrt{gD}$, $z'=\left(z-z_{0}\right)/D$,
and $H'=\left(H_{p}-z_{0}\right)/D$. Bagnold's constitutive equation
\cite{Silbert_etal_2001} states that the pressure $P(z)$ relates
to the shear rate with $\frac{\partial V'_{x}}{\partial z'}\propto\sqrt{P(z')}$.
Integrating this relation and assuming that it holds in a reference
frame moving with velocity $V_{s}$, we shall have $V_{x}'=A(\theta)\times\left(H'^{3/2}-(H'-z')^{3/2}\right)$,
with $A(\theta)$ a constant that is related to the inertial number
defined in the next section. Fig. \ref{fig:Bagnold_collapse} shows
that this is indeed the case, that all the vertical profiles of the
velocity indeed collapse on the theoretical curve above the rolling
layer. The ``sliding velocities'' $V_{s}$ are shown in inset of
Fig. \ref{fig:Bagnold_collapse}, with an excellent fit between the
measured $V_{s}$ at $z_{0}=1.5D$ and the fitted value from the Bagnold
profile. $V_{s}$ increase roughly linearly with $\tan\theta$.

Globally, flows on flat frictional surfaces can thus be decomposed
into a rolling basal layer, above which the main bulk of the flow
follows the classical Bagnold scaling. This is consistent with the
observations reported in \cite{Delannay2007,Taberlet_2005}. Note
that, as we have averaged over the transverse direction, this analysis
however tells nothing on the internal flow structure visible in the
previous sections.

\begin{figure}
\begin{centering}
\includegraphics[width=1\columnwidth]{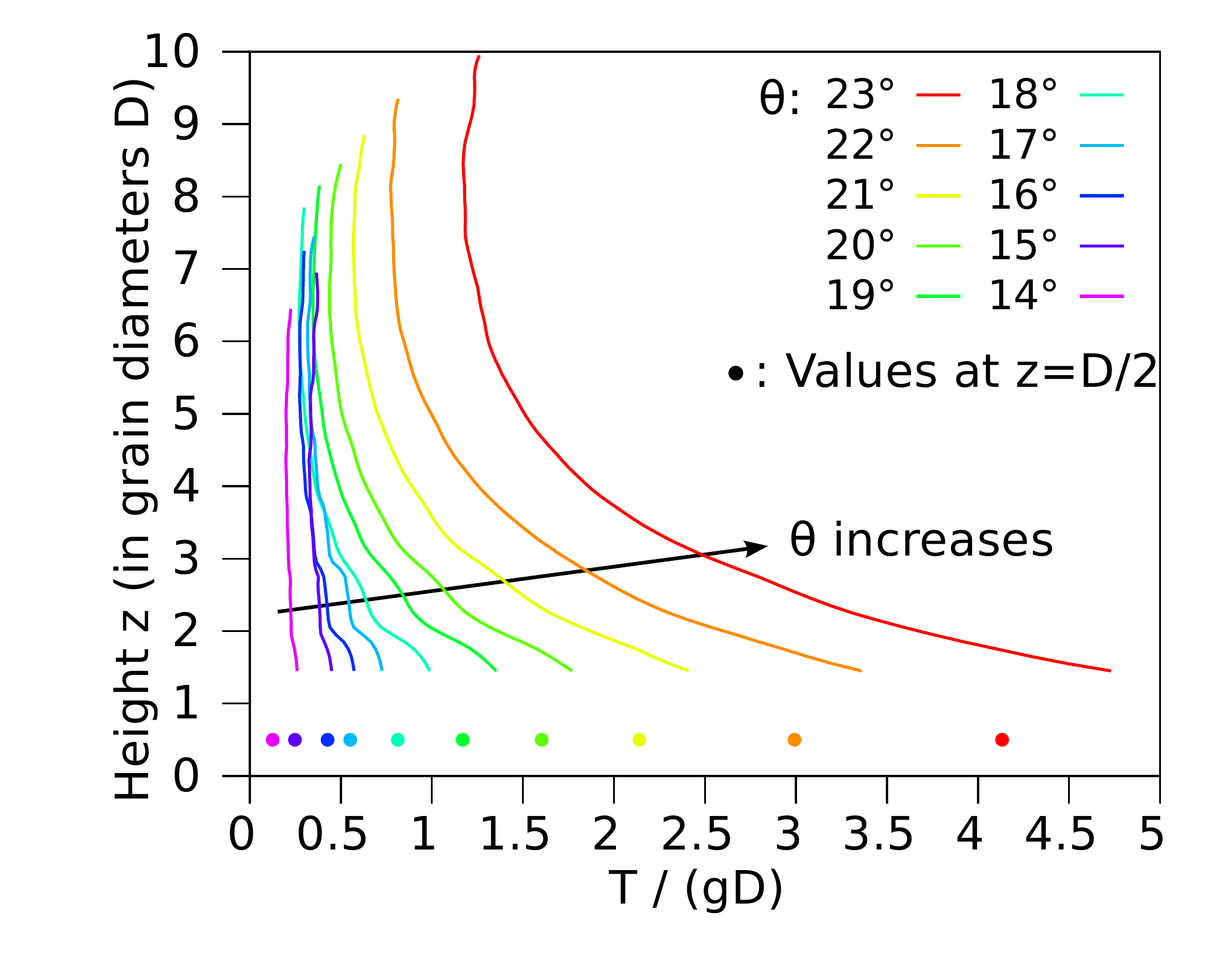} 
\par\end{centering}

\caption{\label{fig:temperatureZ}(Color online). Vertical profile of the velocity
fluctuations, averaged over $y$.}
\end{figure}

Fig. \ref{fig:Flow_and_T} shows the color-coded map of the ``granular
temperature'' $T$ in the cross-section $yz$ plane. Height profiles
of $T$, averaged on the whole width for each $z$, are shown in Fig.
\ref{fig:temperatureZ}, computed in the bulk of the flow above $z_{0}=1.5D$.
The strong velocity gradient at the transition between the bulk and
the basal rolling layer (see Fig. \ref{fig:vprofileZ}) prevents a
meaningful computation of $T$ in that transition. Fig. \ref{fig:temperatureZ}
also shows as separate points the basal ``temperature'' value that
takes into account only contributions from the rolling layer. The
main bulk of the flow thus rests on top of an effective base with
higher ``granular temperature'' and large gradient, which then decreases
according to a height profile compatible with these found in the core
flow on bumpy bases, in numerical simulations \cite{Silbert_etal_2001}
of thick flows. In these simulations~\cite{Silbert_etal_2001}, the
influence of a bumpy base extends to $z=5D$ at which point the ``granular
temperature'' is maximal (Fig. 6 of \cite{Silbert_etal_2001}), and
then it decreases with the height. Note that the temperature profiles
reported in Fig. 15 are also compatible with those predicted by the
kinetic theory~\cite{Forterre_Pouliquen_2002,Ahn_etal_1992} and
with the early 2D simulation results in~\cite{Campbell_Brennen_1985}. 

\begin{figure}
\begin{centering}
\includegraphics[width=1\columnwidth]{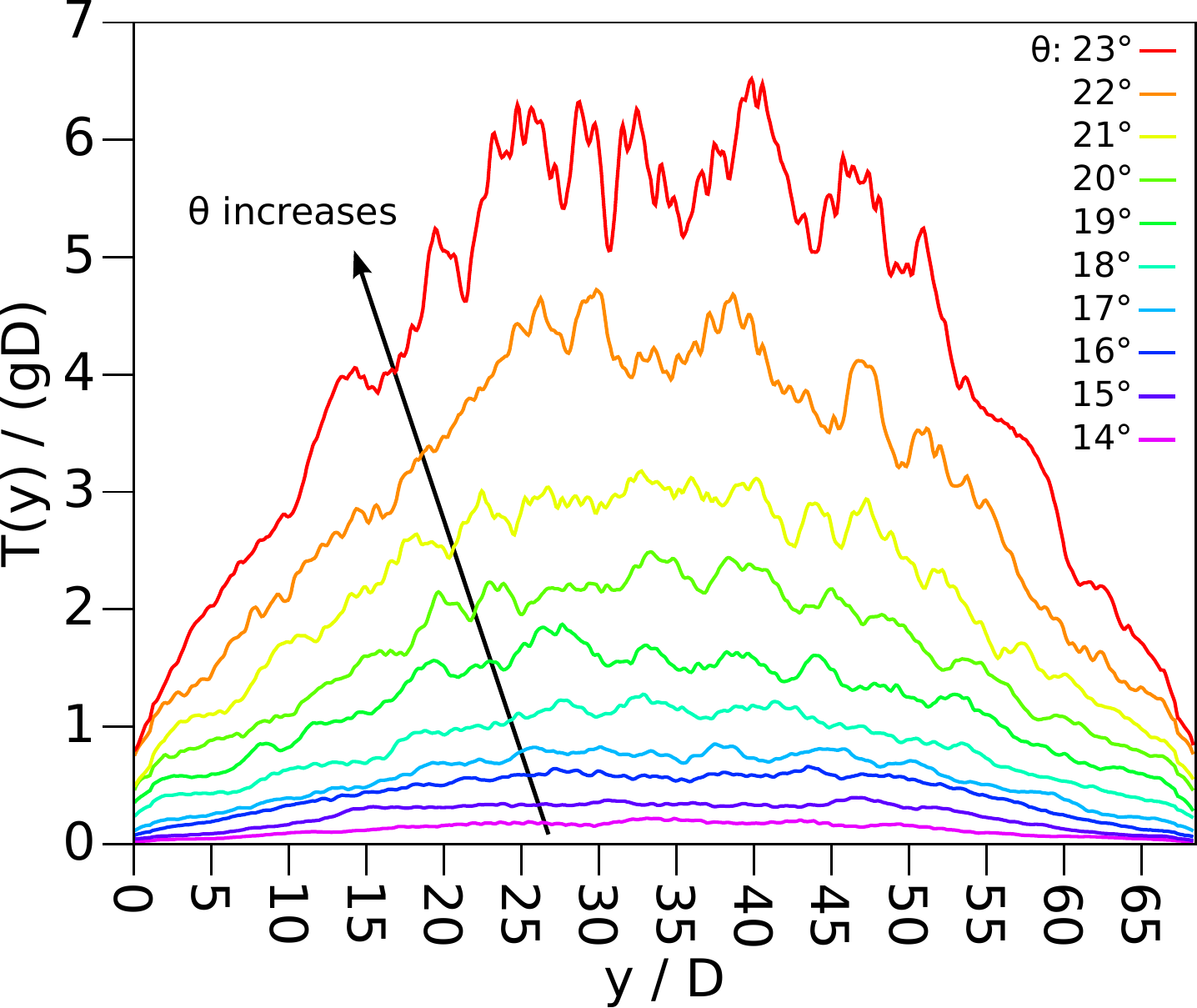} 
\par\end{centering}

\caption{\label{fig:temperatureY}(Color online). Transverse profile of the
velocity fluctuations at the flat base, smoothed over $\pm2.5D$ in
$y$.}
\end{figure}

\begin{figure*}
\begin{centering}
\includegraphics[width=1\textwidth]{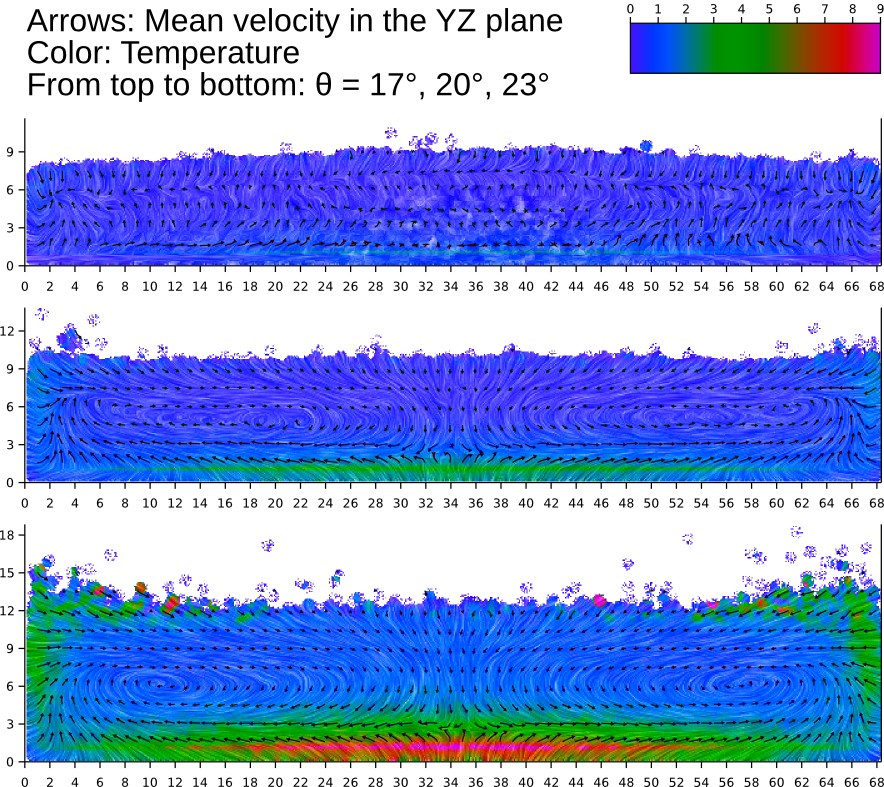} 
\par\end{centering}

\caption{\label{fig:Flow_and_T}(Color online). Vector field of the velocities
in the transverse $yz$ plane on top of the color/gray-coded ``granular
temperature'' $T$. The data, obtained in the steady and fully developed
regime, have been averaged over time and over the periodic cell in
the direction of the flow.}
\end{figure*}

The granular temperature $T$ is nearly constant over the bulk (Fig.
\ref{fig:temperatureZ}) in the unidirectional regime. In the convective
regime, Fig. \ref{fig:Flow_and_T} shows a $z$ profile inversion
between the values at walls and the center that resembles Figs. 15a
and 15b of \cite{Haynes_Walton_2000}, where similar profiles were
computed on a bumpy base with flat frictional walls. The highest temperatures
occur near the center of the base (Fig. \ref{fig:temperatureY}),
the temperature gradient is large near the side walls (at least at
the base). All these features are thus coherent with the idea of a
basal layer producing an effective bumpy base.

\subsection{\label{sub:Rayleigh-Benard-Analogy}Analogy with Rayleigh-Bénard
convection}

Matching experimental evidence for secondary flows was first seen
in \cite{Savage1979}, on a bumpy base. Spontaneous generation of
longitudinal vortices in rapid granular flows down rough inclined
planes are also been reported in \cite{Forterre_Pouliquen_2001}.
The dense and faster troughs correspond to the downward part of the
flow, while the dilute and slower crest correspond to the upward part.
In order to explain them an analogy with Rayleigh-Bénard convection
was proposed \cite{Forterre_Pouliquen_2002} for granular flows based
on considerations from the kinetic theory for granular gases \cite{Jenkins_Savage_1983}.
A three-dimensional linear stability analysis of SFD flows reveals
that in a wide range of parameters, they are unstable under transverse
perturbations. The structure of the unstable modes is globally in
good agreement with the rolls we observe in the main plug of our flows
on a flat base, despite packing fractions reaching high values $\nu_{max}>0.4$
in the core in our case (see Fig. \ref{fig:PackingFractionZ}), unlike
the experiments on a rough base reported in \cite{Forterre_Pouliquen_2002}
where $\nu_{max}\approx0.2$. These values of $\nu_{max}$ are similar
to those obtained in the numerical simulations of \cite{Borzsonyi_etal_2009}.
In \cite{Borzsonyi_etal_2009} two different regimes of stripes are
described, but the granular temperature is not available. The ``dilute''
regime corresponds to the regime described in \cite{Forterre_Pouliquen_2001,Forterre_Pouliquen_2002}
where the dense fast region with downwards motion corresponds to a
height minimum, while in the ``dense'' regime it corresponds to
a height maximum. The dense regime is observed for an average packing
fraction $\overline{\nu}$ comprised between 0.36 and 0.57, while
$0.12<\overline{\nu}<0.42$ in the dilute regime. It is difficult
in our case to know which regime correspond our rolls correspond to.
The average density is in the common range, and the curvature of the
surface is not a clear indication as the walls could deform it.

Fig. \ref{fig:Flow_and_T} shows the color-coded map of $T$ and the
velocity field in the cross-section $yz$ plane. We can see that the
motion in the bulk part of the transverse plane consists of a pair
of counter-rotating vortices - The Fig. 5b in \cite{Savage1979} shows
a similar roll orientation. The material moving towards the base,
in the central part is flowing faster in $x$ direction than the grains
rising up on the sides. The average density is higher where the flow
is downwards and smaller where the flow is upwards. We also observe
the temperature vertical profile inversion reported in Fig. 8d of
\cite{Forterre_Pouliquen_2002}: the temperature gradient is opposed
to the transverse velocity in the downward and upward parts of the
vortices.

Interestingly, the Rayleigh-Bénard regime is similar to the convection
that occurs when a granular bed on a bumpy base is shaken at high
intensity~\cite{Eshuis_PRL_2010}. In such a system the shaking and
the bumpiness of the base lead to a higher granular temperature in
the vicinity of the base. The granular bed is then heated from below
and cooled from above. In our system, as shown above, the granular
layer in contact with the flat frictional base can be considered as
a bumpy sliding base atop which a sheared flow occurs.

\section{\label{sec:Viscoplastic-rheology}Viscoplastic rheology}

A viscoplastic rheology for incompressible flows was proposed in \cite{Jop_etal_2006},
as a 3D extension of the proposal in \cite{GDRMIDI2004}. We expect
it to hold in the unidirectional flows case, for which \cite{GDRMIDI2004}
was proposed. In \cite{Borzsonyi_etal_2009}, Börzsönyi \textit{et
al.} have shown that the viscoplastic rheology does not hold locally
for granular convection in the bumpy boundaries case. They then propose
an extension for this rheology to compressible flows.

This section investigates the situation for the flat frictional scenario,
with an effective basal layer, together with a form of the rheology
using averaged quantities for a global analysis. We average the constitutive
equation proposed in \cite{Jop_etal_2006} to match the vertical profile
of $P(z)$ presented in the previous section, leading to the definition
of an inertial number $I(z)=\left|\left\langle \dot{\gamma}\right\rangle (z)\right|D/\sqrt{P(z)/\rho_{g}}$.
In this expression the strain rate tensor $\dot{\gamma}$ is averaged
at each $z$ location over $L$ and $W$, $\left\langle \dot{\gamma}\right\rangle (z)=\frac{1}{LW}\int_{x=0}^{L}\int_{y=0}^{W}\dot{\gamma}dxdy$,
and the norm used is the same as in \cite{Jop_etal_2006} (i.e. $\left|a\right|=\sqrt{\frac{1}{2}\sum_{i,j}a_{ij}^{2}}$,
which recovers the 1D expression of $I$ from \cite{GDRMIDI2004}
when the $\dot{\gamma}$ tensor is strongly dominated by $\frac{\partial v_{x}}{\partial z}$).
We use the definition of $I$ from \cite{Jop_etal_2006} in which
the grain density $\rho_{g}$ is used for normalization, while the
original proposal \cite{GDRMIDI2004} used the density of the continuum
$\rho_{c}=\rho_{g}\bar{\nu}$. Literature on this topic shows that
both approaches are in use (\textit{e.g.} \cite{Cruz_etal_2005,Jop_etal_2005}
use $\rho_{g}$, \cite{Borzsonyi_etal_2009} use $\rho_{c}$). The
averaging we propose cancels the transverse $v_{y}$ and vertical
$v_{z}$ velocity components thanks to the symmetry of the inner rolls
(see Fig. \ref{fig:Flow_and_T}), recovering an expression that effectively
behaves as for a unidirectional flow without internal structure. Moreover,
we checked that the norm of $\left\langle \dot{\gamma}\right\rangle (z)$
differs from the norm of its deviator by less than $0.35$\%, hence
the condition for an incompressible flow is satisfied on average (i.e.
local dilations that may occur along $y$ in the rolls, if any, cancel
in any given $z$ slice). In these conditions, we expect the constitutive
equation using the average $\left\langle I\right\rangle $ to hold
quite well, which is indeed the case (see below).

\begin{figure}
\begin{centering}
\includegraphics[width=1\columnwidth]{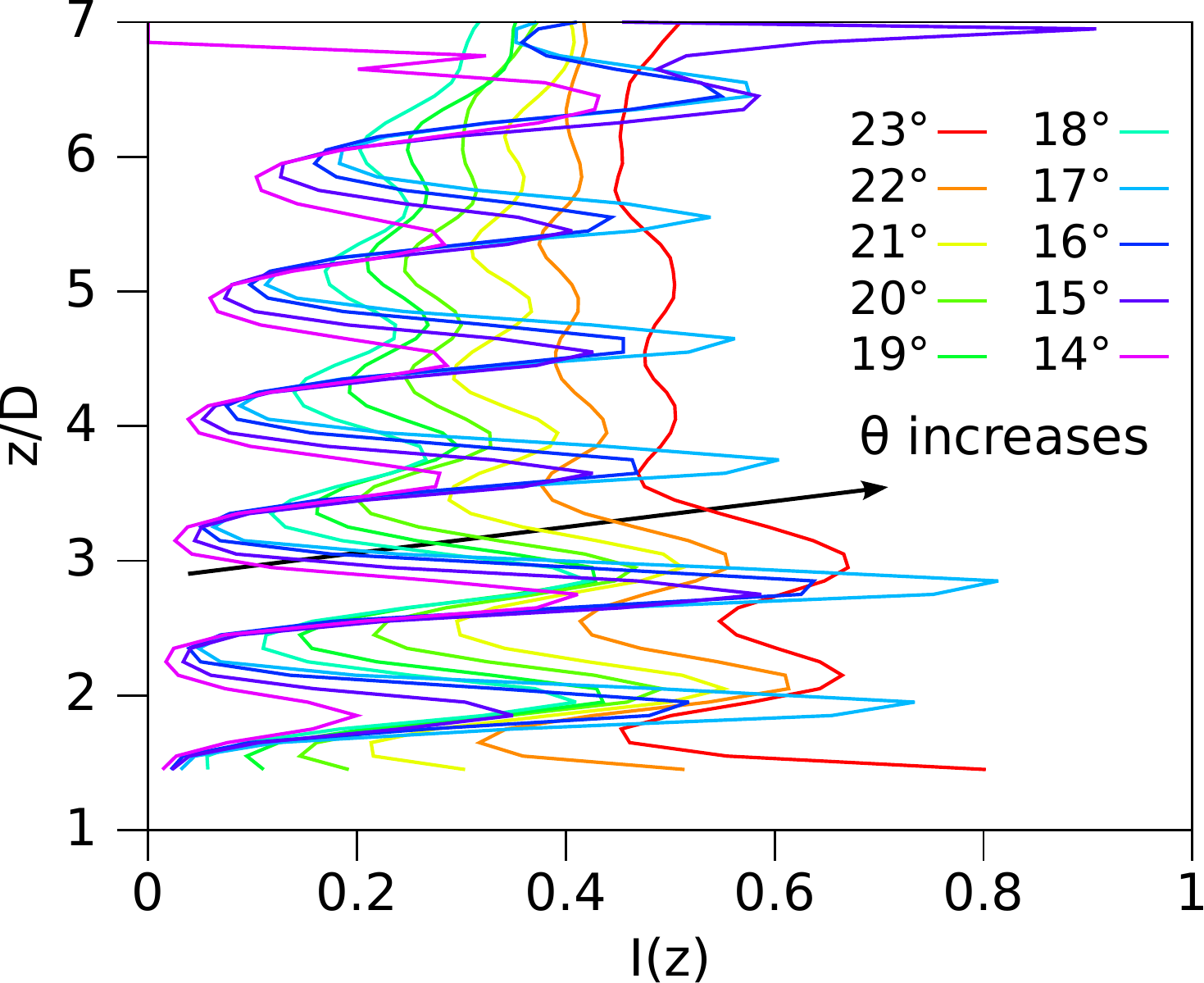} 
\par\end{centering}

\caption{\label{fig:I(z)}(Color online). Vertical profile of the inertial
number $I$.}
\end{figure}

Fig. \ref{fig:I(z)} shows the vertical profile of $I(z)$. Above
the basal layer the average $\left\langle I(z)\right\rangle _{6.5>z/D>z_{0}/D=1.5}$
is defined in the main bulk of the flow. Oscillations over that averaged
value are present in the unidirectional regimes, matching the layered
structure previously mentioned in Section \ref{sub:Packing-fraction},
while for the convective regimes the vertical profile does not vary
as much $I(z)\approx\left\langle I\right\rangle $. Note that for
the dense flows on bumpy base case, $I$ is assumed to be constant
on the whole height (Section 8.4.1 of \cite{GDRMIDI2004}).

\begin{figure}
\begin{centering}
\includegraphics[width=1\columnwidth]{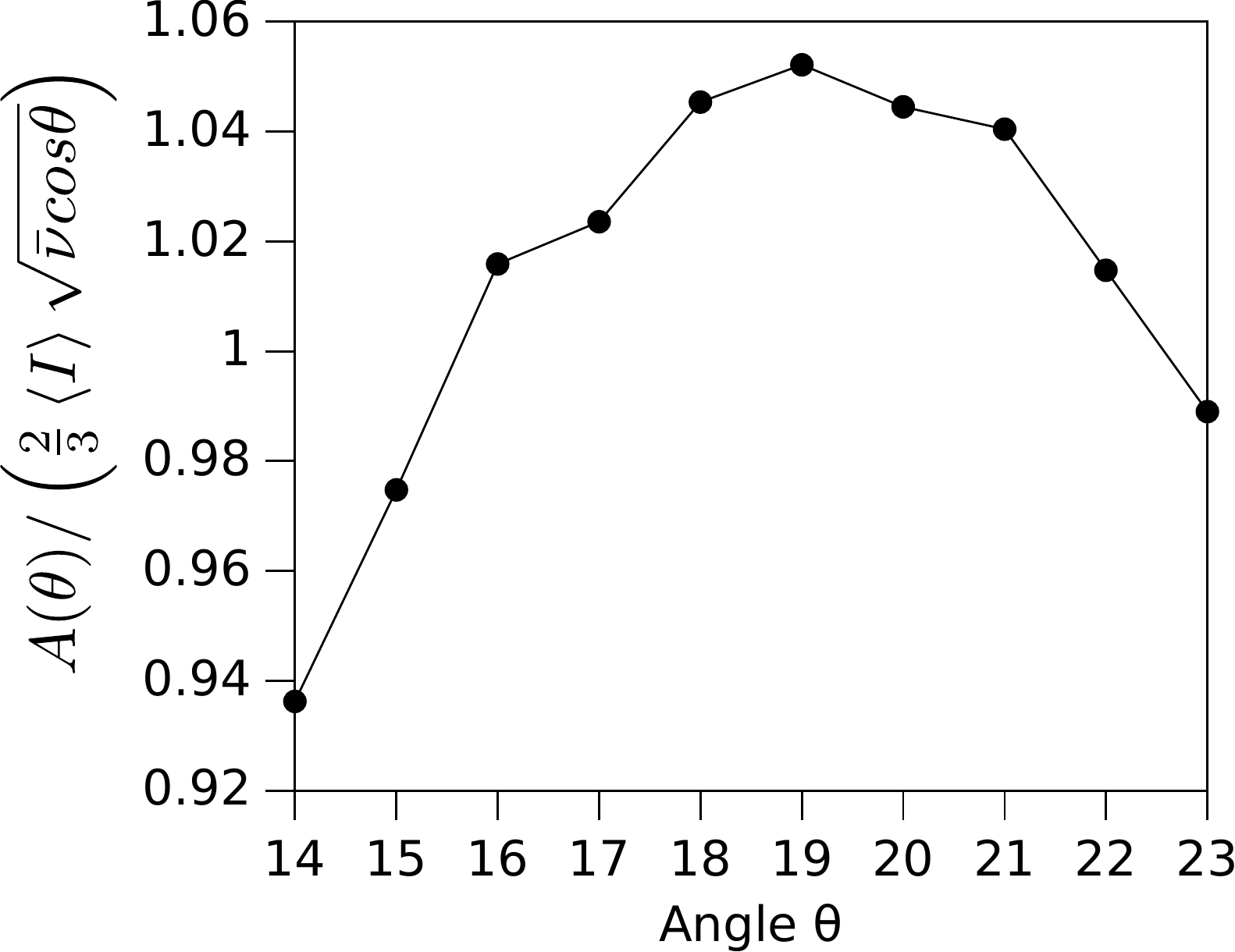} 
\par\end{centering}

\caption{\label{fig:A/I}Fit quality of the theoretical Bagnold profile for
each angle.}
\end{figure}

If the constitutive equation of \cite{Jop_etal_2006} holds in the
averaged form we propose, we shall recover a velocity profile in the
form of a Bagnold scaling (eq. 25 of \cite{GDRMIDI2004}), with $A(\theta)=\frac{2}{3}\left\langle I\right\rangle \sqrt{\bar{\nu}cos\theta}$
matching the constant fitted in the previous section. Fig. \ref{fig:A/I}
shows that this is indeed the case, up to a worst-case 5\% accuracy.

\begin{figure}
\begin{centering}
\includegraphics[width=1\columnwidth]{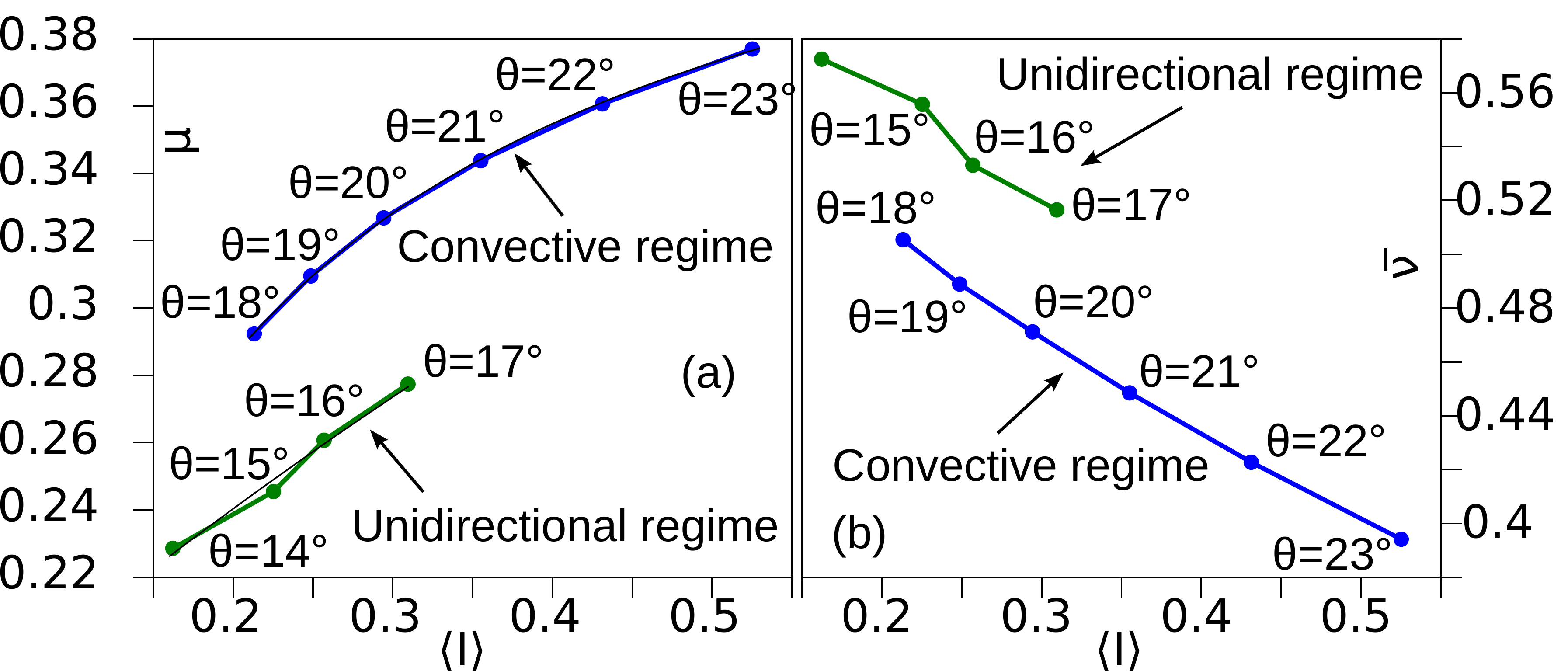} 
\par\end{centering}

\caption{\label{fig:muI}(Color online). (a) Effective friction $\mu$ and
(b) average packing fraction $\bar{\nu}$ vs average inertial number
$\left\langle I\right\rangle $ in the bulk of the flow. Overlaid
thin black lines in (a) are the empirical fits mentioned in the main
text.}
\end{figure}

From the momentum balance for the main bulk flowing on the basal layer,
including wall effects, it is possible to obtain $\mu(\left\langle I\right\rangle )$,
the effective friction coefficient of the main bulk on the basal layer:
$\mu=tan(\theta)-\widehat{\mu_{W}}\left(H_{p}-z_{0}\right)/W$ \cite{Jop_etal_2005,Taberlet2003}.
When $\left\langle I\right\rangle $ is plotted against $\mu(\left\langle I\right\rangle )$
the separation between the unidirectional and the convective regimes
is clearly apparent as a discontinuity, see Fig. \ref{fig:muI}a.
The best fit parameters for the constitutive equation $\mu(\left\langle I\right\rangle )=\mu_{s}+\left(\mu_{2}-\mu_{s}\right)/\left(1+I_{0}/\left\langle I\right\rangle \right)$
proposed in \cite{Jop_etal_2006} (see Fig. \ref{fig:muI}a) are $\mu_{s}=0.0046$,
$\mu_{2}=0.479$ and $I_{0}=0.13$ in the convective regime, while
$\mu_{s}=0.16$, $\mu_{2}=1.02$ and $I_{0}=2.06$ in the unidirectional
regime, showing that the empirical constitutive equation $\mu(\left\langle I\right\rangle )$
changes during the transition. The break is similarly visible on the
$\bar{\nu}$ vs $\left\langle I\right\rangle $ profile in Fig. \ref{fig:muI}b.
Both branches decrease nearly linearly, compatible with Fig. 2 in
\cite{Cruz_etal_2005} where the model coefficient of restitution
and spring stiffness are varied in a 2D simulation, and unlike Fig.
4e of \cite{Borzsonyi_etal_2009} where the $\bar{\nu}(I)$ relation
is built locally and not in averaged form.

The number $I$ can also be interpreted as the ratio of a macroscopic
rearrangement time scale over a shearing time scale \cite{GDRMIDI2004}.
The observed drastic reduction in $I$ at the transition between the
unidirectional and convective regimes reflects the fact that granular
convection rearranges the grains much faster than slow diffusion within
the ordered layering. The $\left\langle I\right\rangle $ value at
$\theta=18$\textdegree{} in Fig. \ref{fig:muI} is consistent with
the convective regime despite computations being performed in the
oscillating state, leading to the hypothesis that the oscillations
are related to the onset of convection. That hypothesis will be investigated
in a future work.

\section{Conclusion}

Our numerical simulations with side walls generate SFD flows comparable
to the experimental setup \cite{Louge_Keast_2001} with a compatible
range of angles, distances of establishment and velocity profiles.
We confirm that the influence of the friction on the lateral walls
is negligible (\cite{Louge_Keast_2001} and Section \ref{sec:Viscoplastic-rheology}),
but also that walls manifest in other ways a long-range influence
within the flow (\cite{Louge_Keast_2001} and Fig. \ref{fig:vprofileY}).
In any case, side walls cannot be ignored even when they are far away,
especially since channeled flows can be directly compared to experiments.
Building on these results we extrapolate the simulation to larger
inclination angles and find that distances for reaching the steady
states exceed the experimental chute length. These regimes also correspond
to the presence of granular convection, whereby grains are circulated
within the whole flow, unlike the unidirectional regimes where grains
mostly remain in a ``crystallized'' layered structure.

Compared to the well-studied bumpy base scenario, flows on flat frictional
surfaces involve a much faster overall velocity, thanks to the presence
of a basal layer of rolling grains, upon which slides the main bulk
of the flow. We then interpret that bottommost layer of grains as
an effective base for the flow bulk and we show that in these conditions,
the bulk follows a conventional Bagnold scaling. The analogy with
an effective rough base extends to the presence of a convective regime
with similar velocity and density profiles. However, due to the increased
overall velocity, and owing to the effective base being less rigid
than a fixed bumpy one, the convection rolls appear for lower angles
and mass holdups in the flat frictional case than in the bumpy one.

As for the bumpy case, we find that over the effective base the bulk
of the flow follows on average a viscoplastic rheology \cite{Jop_etal_2006},
for each of the SFD regimes. The transition between these regimes
corresponds to a break in the friction $\mu$ versus inertial number
$I$ relation (Fig. \ref{fig:muI}), with a drastic reduction in $I$
that matches the effect of the secondary rolls (faster grain rearrangement).

Channeled flows down flat frictional surfaces are well adapted for
testing granular rheologies numerically and studying boundary conditions.

\paragraph*{Acknowledgments}

We thank Région Bretagne for funding (CREATE Sampleo Grant), M.Y.Louge,
and J.T.Jenkins for helpful discussions and comments. We are grateful
to Sean McNamara for a careful reading of the manuscript.

\bibliographystyle{apsrev4-1}
\bibliography{convection_lisse}

\end{document}